\documentclass{aa}  
\usepackage{graphicx}
\usepackage{txfonts}
\usepackage{soul}
\usepackage[colorlinks=true, citecolor=blue]{hyperref}
\usepackage{natbib,twoopt}
\bibpunct{(}{)}{;}{a}{}{,}             %% natbib format for A&A and ApJ
\usepackage{hyperref}
\hypersetup{
    colorlinks=true,
    linkcolor=blue,
    citecolor = blue,
    filecolor=magenta,
    urlcolor=blue}
\usepackage{booktabs}

\begin{document}

\authorrunning{Ferraro et al.}
\titlerunning{The Bulge Cluster Origin (BulCO) survey}

\title{The Bulge Cluster Origin (BulCO) survey at the ESO-VLT: probing the early history of the Milky Way assembling. Design and first results in Liller~1\thanks{Based on observations collected at the Very Large Telescope of the European Southern Observatory at Cerro Paranal (Chile) under Large Program 110.24A4 (PI:Ferraro)}}

\author{F. R. Ferraro\inst{1}\fnmsep\inst{2}, L. Chiappino\inst{1}\fnmsep\inst{2}, A. Bartolomei\inst{1,2}, L. Origlia\inst{2}, C. Fanelli\inst{2}, B. Lanzoni\inst{1}\fnmsep\inst{2}, C. Pallanca\inst{1}\fnmsep\inst{2}, M. Loriga\inst{1}\fnmsep\inst{2}, S. Leanza\inst{1}\fnmsep\inst{2},   E. Valenti\inst{3,4}, D. Romano\inst{2}, A. Mucciarelli\inst{1}\fnmsep\inst{2}, D. Massari\inst{2}, M. Cadelano\inst{1}\fnmsep\inst{2}, E. Dalessandro\inst{2}, C. Crociati\inst{5} and R.M. Rich\inst{6}
    }

   \institute{Dipartimento di Fisica e Astronomia, Università degli Studi di Bologna, Via Gobetti 93/2, I-40129 Bologna, Italy  \email{francesco.ferraro3@unibo.it}
   \and
   INAF, Osservatorio di Astrofisica e Scienza dello Spazio di Bologna, Via Gobetti 93/3, I-40129 Bologna, Italy
         \and
         European Southern Observatory, Karl-Schwarzschild-Strasse 2, 85748 Garching bei Munchen, Germany
         \and
Excellence Cluster ORIGINS, Boltzmann-Strasse 2, D-85748 Garching Bei Munchen, Germany
\and
Institute for Astronomy, University of Edinburgh, Royal Observatory, Blackford Hill, Edinburgh, EH93HJ, UK
\and
         Department of Physics and Astronomy, UCLA, 430 Portola Plaza, Box 951547, Los Angeles, CA 90095-1547, USA\\
         }

\abstract
{We present the scientific goals and the very first results of the Bulge Cluster Origin (BulCO) survey. This survey has been  specifically designed to perform an unprecedented chemical screening of  stellar systems orbiting the Milky Way bulge, with the aim to unveil their true origin. It takes advantage of the improved performances of the spectrograph CRIRES+ operating at the ESO Very Large telescope, in the near-infrared domain. Due to the complex evolutionary history of the Milky Way, a variety of relics tracing different phenomena is expected to
populate the Bulge: 
globular clusters formed in-situ or accreted from outside the Galaxy,  nuclear star clusters of cannibalized structures, and possibly a few remnants of the proto-bulge formation process (the so-called ``bulge fossil fragments"). The signatures of the different origins are imprinted in the chemical properties of these stellar systems because specific abundance patterns provide authentic “chemical DNA tests” univocally tracing the enrichment process and, therefore, the environment where the stellar population formed. Thus, each system can provide a new piece of information on the bulge formation and evolutionary history. As first results of the survey, here we discuss the $\alpha-$element and iron abundances of a sample of stars observed in the stellar system Liller 1,  which is proposed to be a bulge fossil fragment. By combining this dataset with a recently published sample of high/mid-resolution spectra, we discuss the overall chemical properties of the stellar populations in Liller1, proving its link with the Galactic bulge and providing new constraints on its star formation history.  
}
\maketitle
\section{Introduction} 
\label{intro}
The $\Lambda$-Cold Dark Matter cosmological model \citep[e.g.][]{white_rees1978, davis85} predicts that cosmic structures assemble bottom-up through hierarchical merging processes, and it successfully explains the existence and the observed properties of massive galaxies and galaxy clusters. Indeed, the unprecedented exploration of the Milky Way (MW) performed by the Gaia mission identified several kinematical structures (as Gaia-Enceladus, Helmi stream, Sequoia, etc.) associated to past accretion events suffered by the MW halo \citep[see, e.g.][]{helmi20}, suggesting that only $\sim50\%$ of the surveyed globular clusters (GCs) has an in-situ origin, while the remaining fraction has been accreted from outside or still have unclear nature \citep{massari19}. In this respect, the most famous remnant of a remote merging event found in the Galactic halo is $\omega$ Centauri. 
 
Despite its original classification of GC, this stellar system has been found to host multi-iron sub-populations \citep{norris+96, pancino+00, origlia+03, ferraro+04, johnson2010_omega, bellini+17, meszaros2021_omega, deimer2022_omega, deimer2024_omega}, and its properties suggest that it is the remnant of a nuclear star cluster of an accreted dwarf galaxy \citep{bekki+03, romano+07}. 
Because of its prohibitive density and reddening conditions, no comparably conclusive results are available for the Galactic bulge. However, it is reasonable to expect that also this region of the Galaxy experienced past accretion events, which left detectable signatures imprinted in its population of old stellar systems. Actually, according to what observed at high-redshift in the so called ``chain and clumpy" galaxies \citep{elme09, genzel11, tacchella15} the entire assembly process of a galaxy starts from the bulge through the merging of local clumps of stars and gas \citep{immeli_04, elme08, bournaud09}. Although most of those massive clumps are predicted to dissolve and form the spheroid, a few of them could survive the total disruption \citep{bournaud16} and be still present in the inner regions of the host galaxy, grossly appearing as massive GCs. At odds with genuine GCs, however, these fossil relics are expected to host multi-iron and multi-age sub-populations, sharing the same chemical patterns drawn by bulge field stars (see below). Two stellar systems with such properties were recently identified in the MW bulge, thus promising to be ``bulge fossil fragments (BFFs)", i.e., the fossil records of its hierarchical assembly process: Terzan 5 \citep{Ferraro_09, ferraro_16, Origlia_11, Origlia_13, origlia_19, Massari_14, romano23, crociati+24} and Liller 1 \citep{Ferraro_09, ferraro_21, pallanca_21, crociati_23, deimer24,fanelli+24}. 
In addition to BFFs, tracing the very first phases of the bulge formation process, and in-situ formed genuine GCs, a variety other objects 
mapping different phenomena are expected to populate the bulge under the appearance of old stellar systems: 
accreted genuine GCs, formed in an external galaxy and brought into the bulge by an accretion event, and, possibly, 
nuclear star clusters of cannibalized structures. The signatures of different origins are imprinted into the physical and chemical properties of these stellar systems. In this framework, we are leading a project aimed at the full characterization of bulge stellar systems, by combining high-resolution and multi-wavelength photometric and spectroscopic observations, proper motion (PM) membership selection, and accurate correction for the effects of differential reddening (see \citealt{massari12, cadelano23, deras23, libralato22, pallanca_21,pallanca21b, pallanca23, saracino_15, saracino16,saracino19}). 
 
The BulCO survey is part of this comprehensive effort. 
Taking advantage of the pioneering experience of our group in near-infrared (NIR) spectroscopy of dense star clusters (see \citealt{origlia97, origlia_02, origlia+03, origlia05, origlia08, origlia04}), and the improved performances of the CRyogenic high-resolution InfraRed Echelle Spectrograph \citep[CRIRES+;][]{kaufl+04, dorn+14, dorn+23} at the ESO Very Large Telescope (VLT), the BulCO survey is aimed to perform an unprecedented high-resolution chemical screening of the stellar populations hosted in a selected sample of systems traditionally cataloged as bulge GCs \citep[e.g.,][]{harris96}. Chemical tagging is a key tool for the determination of the nature and the origin of the stellar systems orbiting the Galactic Bulge. In fact, the characteristic dynamical timescales in the Bulge are so short that dynamical friction and other processes could have cleaned-out the kinematic signatures imprinted into the original orbit of each system, while the chemical properties permanently flag the cluster origin independently of the accretion epoch. The abundances of iron, iron-peak, CNO and other $\alpha$- and light-elements will be measured with the final goal to disentangle the true nature and origin of each target, thus reconstructing the bulge evolutionary history, by means of ``chemical DNA tests". In fact, the atmospheres of the stars that we observe today preserve memory of the chemical composition of the interstellar medium (ISM) from which they formed, and the chemical abundances of the ISM vary in time if more than one burst of star formation occurs, owing to the ejecta of each stellar generation. Thus, stars formed at different times and in environments with different star formation rates (SFRs) have different chemical compositions, and by analyzing the chemistry of each stellar population one can univocally trace the enrichment process of the ISM. Different abundance patterns are expected depending on the stellar polluters, the enrichment timescale and the SFR, with a few specific abundance patterns being so distinctive that they can be used as ``DNA tests" of the stellar population origin. More specifically, (1) BFFs are expected to host multi-iron and multi-age sub-populations with the same chemical patterns drawn by bulge stars, (2) nuclear star clusters of accreted galaxies should show iron spread \citep[e.g.,][]{neumayer2020}, together with $\alpha-$element patterns and iron-peak element abundances substantially different from the Bulge environment, and (3) genuine GCs are predicted to be homogeneous in iron and show peculiar correlations and anti-correlations between pairs of light-elements \citep[e.g][]{carretta09, carretta10, milone17}; among these, in-situ formed and accreted GCs can be distinguished through their $\alpha$- and iron-peak element abundances \citep{minelli21, mucciarelli21, ceccarelli+24}.
 
The aim of this paper is to present and describe the BulCO survey, and discuss the first set of results for a sample of stars observed in 
Liller 1, one of the two candidate BFFs known so far.  Section \ref{bulco} provides an overview of the survey. In Section 3, we describe the observations and the adopted data reduction procedures. The results are presented in Section 4, while Section 5 is devoted to the discussion and conclusions of the work.

 \begin{table}
    \scriptsize
    \renewcommand{\arraystretch}{1.25}
    \setlength{\tabcolsep}{13pt}
    \caption{Main properties of the 17 star clusters selected as targets of the BulCO survey. }
    \begin{tabular}{|c|c|c|c|c|}
    \hline\hline
    Cluster Name &  $M_V$   &  E(B-V)  &   [Fe/H]    &   Core radius    \\
    \hline
       & [mag] & [mag] & [dex]  & [arc min]  \\
    \hline
    NGC 6304   &  -7.3 &  0.5 & -0.45 & 0.21   \\
    NGC 6316  &  -8.3  &  0.5  & -0.45 & 0.65   \\
    NGC 6380   &  -7.5  &  1.2  & -0.75 & 0.34  \\
    NGC 6388   &  -9.4 &  0.4  & -0.55 & 0.12   \\
    NGC 6441  & -9.6   &  0.5  & -0.46 & 0.13  \\
    NGC 6528  & -6.6   & 0.5  & -0.11 & 0.13  \\
    NGC 6553  & -7.8   & 0.6  & -0.18 & 0.53  \\
    NGC 6656  &  -8.5 &  0.3 & -1.70 & 1.33  \\
    Palomar6  & -6.8   &  1.5  & -0.91 & 0.66  \\
    Terzan 1  &  -4.4  &  2.0 & -1.03 & 0.04  \\
    Terzan 2  & -5.9   & 1.9  & -0.69 & 0.03  \\
    Terzan 5  & -7.4   & 2.3  & -0.23 & 0.16   \\
    Terzan 6  &  -7.6  & 2.3  & -0.56 & 0.05  \\
    Terzan 10 & -6.3   &  2.4 & -1.00 & 0.90  \\
    UKS1      & -6.9  &  3.1  & -0.64 & 0.15  \\
    Djorgovski 2  & -7.0 &  0.9    & -0.65 & 0.33  \\
    Liller 1   & -7.3  & 4.5   & -0.33 & 0.06  \\
    \hline\hline
    \end{tabular}
    \label{tab1}   
\vspace{0.15cm}
Notes: \\
All values are from the \citet{harris96} catalog, except for the color excess of Liller 1 that is from \citet{pallanca_21}
    \end{table}

\begin{figure}
    \centering
    \includegraphics[width=\columnwidth]{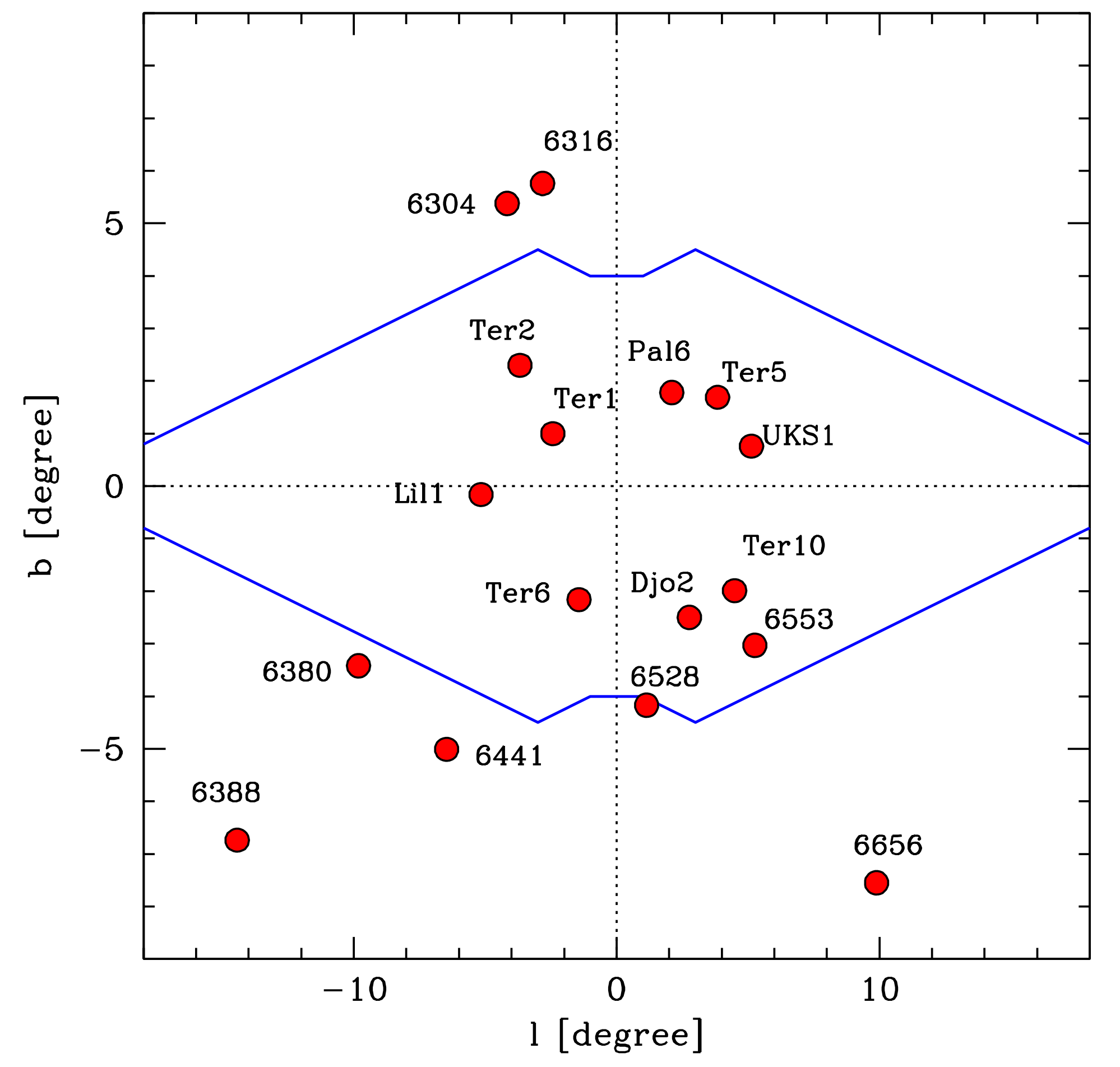}
    \caption{Position in Galactic longitude (l) and latitude (b) of the 17 star clusters (large red circles) selected as targets of the BulCO survey. They are all observed in the direction of the Galactic bulge. For reference, the blue lines represent the outline of the inner bulge as traced at 3.5 $\mu m$ by COBE/DIRBE \citep{weiland94}. } 
\label{map}
\end{figure}
 
\section{The BulCO survey}
\label{bulco}
The BulCO survey is a 255 hour Large Program at the ESO-VLT, specifically designed to take advantage of the improved performances of the CRIRES+ spectrograph. 
 
{\it The selected clusters $-$} 17 stellar systems traditionally classified as GCs in the MW bulge have been selected as spectroscopic targets for the BulCO survey, with the specific aim of decoding their chemical DNA and unveil their true origin and nature. The main properties of the selected targets are listed in Table \ref{tab1}, while their position in Galactic longitude and latitude is shown in Figure \ref{map}. The list includes star clusters that are currently suspected to belong to different categories: the two most metal-rich and (as far as we know) mono-metallic GCs in the MW (NGC 6553 and NGC 6528), the two most massive bulge stellar systems (NGC 6388 and NGC 6441; based on a few spectra, they are suspected to be accreted systems; \citealt{minelli21}; but see also \citealt{carretta23,massari+23}), NGC 6656 (a largely debated GC, suspected to harbor multi-iron sub-populations; \citealt{marino09, mucciarelli15}), the two currently known BFF candidates (Terzan 5 and Liller 1), and a representative sample of stellar systems in the inner bulge that still remain almost chemically unexplored, some of which showing orbital properties consistent with an accreted origin (\citealp{massari19,callingham+22}; namely, Terzan 1, Terzan 2, Terzan 6, Terzan 10, Palomar 6, UKS1, Djorgovski 2, NGC 6304, NGC 6316, NGC 6380). 

{\it The observational strategy $-$} To select the spectroscopic target stars in each of the 17 clusters, we made use of the large photometric dataset collected by our group over the years both in the NIR and in the optical bands (see \citealt{ferraro00, Ferraro_09, ferraro_21, valenti05, valenti07, pallanca_21, lanzoni07, lanzoni10, lanzoni13}). This was complemented by ongoing HST and GEMINI observations, as well as archive images retrieved from the GEMINI, VLT  and HST repositories. Whenever possible, relative PMs have been computed by using multi-epoch HST images (see \citealt{libralato22}) or by combining HST and adaptive optics assisted ground-based images (see examples in \citealt{ferraro_16, ferraro_21, dalessandro16, dalessandro_22}). We also took advantage of PM measurements from Gaia DR3 \citep{gaia_16,gaia_23}, and accurate radial velocities obtained from the ongoing spectroscopic survey MIKiS \citep{ferraro18} and archive data.
 
\begin{table*}[!t]
\caption{Coordinates, K-band magnitude, atmospheric parameters, RVs and chemical abundances of the observed stars in Liller 1.  }

\label{tab2}
\scriptsize
\setlength{\tabcolsep}{8.1pt}
\renewcommand{\arraystretch}{1.4}
    \begin{tabular}{|c|c|c|c|c|c|c|c|c|c|c|}
    \hline\hline
    ID &  RA   &  Dec  &     K & $T_e$   & log(g) & $R_V$  & [Fe/H]  & [Mg/H] & [Si/H] & [Ca/H] \\
    \hline
       & [Deg] & [Deg] & [mag]  & K  & dex & km/s & dex & dex & dex & dex    \\
    \hline
        100760 &   263.35015940 &   -33.39933860 &     12.485 &       4200 &       1.50 &       68.6 &      -0.41 $\pm$       0.01 (17)&      -0.14 $\pm$       0.02 (3)&      -0.09 $\pm$       0.01 (3)&      -0.16 $\pm$       0.03 (2)\\
    300094 &   263.34081200 &   -33.38945370 &      9.359 &       3400 &       0.25 &       60.7 &      -0.46 $\pm$       0.02 (8)&      -0.18 $\pm$       0.10 (1)&      -0.17 $\pm$       0.10 (1)&      -0.21 $\pm$       0.10 (1)\\
    300315 &   263.34635890 &   -33.38677290 &      9.216 &       3350 &       0.25 &       70.6 &      -0.50 $\pm$       0.02 (7)&      -0.21 $\pm$       0.10 (1)&      -0.23 $\pm$       0.02 (2)&      -0.27 $\pm$       0.10 (1)\\
    300553 &   263.34974320 &   -33.38908650 &     11.774 &       4050 &       1.25 &       68.4 &      -0.42 $\pm$       0.01 (18)&      -0.14 $\pm$       0.02 (3)&      -0.17 $\pm$       0.01 (5)&      -0.16 $\pm$       0.02 (3)\\
    300614 &   263.34576520 &   -33.38735650 &     12.454 &       4150 &       1.75 &       74.7 &       0.27 $\pm$       0.01 (19)&       0.24 $\pm$       0.01 (3)&       0.29 $\pm$       0.02 (5)&       0.26 $\pm$       0.01 (4)\\
    300682 &   263.34970830 &   -33.38889850 &     12.460 &       4150 &       1.50 &       73.6 &      -0.27 $\pm$       0.01 (19)&       0.03 $\pm$       0.01 (3)&       0.06 $\pm$       0.02 (3)&       0.01 $\pm$       0.02 (3)\\
    300727 &   263.34491220 &   -33.38586660 &     12.496 &       4200 &       1.75 &       71.6 &       0.28 $\pm$       0.01 (20)&       0.26 $\pm$       0.02 (4)&       0.28 $\pm$       0.01 (7)&       0.31 $\pm$       0.01 (3)\\
    387099 &   263.34466130 &   -33.38908300 &      8.925 &       3400 &       0.50 &       71.3 &       0.19 $\pm$       0.02 (7)&       0.18 $\pm$       0.10 (1)&       0.23 $\pm$       0.02 (3)&       0.21 $\pm$       0.10 (1)\\
    400065 &   263.35796140 &   -33.38444410 &      9.263 &       3400 &       0.50 &       85.8 &       0.20 $\pm$       0.01 (12)&       0.24 $\pm$       0.10 (1)&       0.23 $\pm$       0.02 (4)&       0.22 $\pm$       0.01 (2)\\
    400778 &   263.35441500 &   -33.38819150 &     12.862 &       4300 &       1.75 &       71.6 &      -0.17 $\pm$       0.01 (16)&       0.08 $\pm$       0.02 (2)&       0.05 $\pm$       0.02 (3)&       0.05 $\pm$       0.02 (4)\\
    400829 &   263.35164550 &   -33.38821260 &     12.801 &       4250 &       1.75 &       64.2 &      -0.30 $\pm$       0.01 (16)&       0.03 $\pm$       0.02 (4)&       0.04 $\pm$       0.01 (6)&       0.01 $\pm$       0.02 (4)\\
    400887 &   263.35383770 &   -33.38732720 &     12.927 &       4300 &       2.00 &       71.3 &       0.26 $\pm$       0.01 (19)&       0.24 $\pm$       0.01 (3)&       0.23 $\pm$       0.02 (4)&       0.25 $\pm$       0.02 (4)\\
\hline\hline
\end{tabular}
\vspace{0.15cm}
Notes: \\
The quoted errors are the standard deviations divided by the square root of the number of lines used (reported in brackets), with the exception of a conservative assumption of 0.1 dex in the case when only one line was measurable.
\end{table*}

This allowed us to select at least 20 giant stars in each cluster, for a grand total of 420 spectroscopic targets, that, based on their PM and radial velocity, are likely members of the systems. All the observations have been planned with the 0.4$\arcsec$ wide slit, thus providing an overall spectral resolution $R\sim 50,000$. For each giant, observations in at least two of the following three gratings have been foreseen: J1226, H1582, K2166. These gratings sample a large number of unblended spectral lines, specifically: 
 
\begin{itemize}
    \item the J1226 grating, covering the 1116-1356 nm spectral region, allows the abundance measure of iron, iron-peak elements as Zinc, Vanadium, Chromium, Cobalt, Nickel, and Manganese, $\alpha$-elements as Calcium, Silicon, Magnesium, Titanium, and other light elements as Carbon, Nitrogen, Sodium, Aluminum, and Potassium;
    \item the H1582 grating (sampling $\Delta\lambda=1484$-$1854$ nm) allows the measure of Oxygen and $^{12}$C+$^{13}$C abundances from a few dozen OH and CO molecular lines, respectively, Sulfur, a few ionized lines of the Cerium, as well as other lines of V, Fe, Cr, Mn, Ni, Co, C, Mg, Si, Ca, Ti, Na, Al, and N (from a few dozens of CN molecular lines);
    \item the K2166 grating  
    (covering $\Delta\lambda=1921$-$2472$ nm) allows the identification of about 12 iron lines, a few dozens of CN molecular lines providing N abundance, and $\sim 25$ lines of $\alpha$-elements such as Mg, Si, Ca, Ti, Na, and Al.    
\end{itemize}

Thus, the selected gratings offer the opportunity to measure all the element abundances and abundance ratios necessary to constrain the cluster origin, enrichment timescales and stellar nucleosynthesis. In this respect, the following chemical diagnostics are especially relevant:
\begin{itemize}
    \item[$\bullet$] The [Fe/H] distribution within each cluster is crucial to immediately discriminate whether the system is a potential BFF (showing multi-iron sub-populations, as in Terzan 5 and Liller 1; see \citealp{Massari_14, Origlia_11, crociati_23, fanelli+24}), or the nuclear star cluster of an accreted galaxy (with a broad iron distribution peaking at significantly lower metallicity, as in the case of $\omega$ Centauri; see \citealp{norris+96, johnson2010_omega}), or a GC (with single-iron population).
    \item[$\bullet$] The [$\alpha$/Fe] abundance ratios provide crucial information on the relative contribution and timescale of ISM enrichment from supernovae type II and type Ia (SNeII and SNeIa, respectively), and they can also provide additional constraints to an in-situ (bulge) versus an external (other Galactic components or extra-Galactic) formation scenario.
    \item[$\bullet$] Also the [iron-peak/Fe] abundance ratios of a few iron-peak elements as Vanadium (V) and Zinc (Zn) have been found to be powerful diagnostics of the SFR in the environment where stars formed. In fact, these elements are mainly produced by high-mass ($>20 M_\odot$) stellar progenitors, via hypernovae, SNeII, and electron-capture SNe \citep[][however, see \citealp{palla2021} about the uncertain contribution to V synthesis by SNeIa]{romano10, koba20}. Hence, the [V/Fe] and [Zn/Fe] ratios are expected to be smaller in galaxies with low SFR (hence, with smaller contribution from massive stars), than in high SFR systems \citep{jera18}, allowing a solid distinction between in-situ formed and accreted GCs \citep{minelli21}.
    \item[$\bullet$] The O, Na, Al, Mg and K abundance distributions within each cluster can probe whether there has been self-enrichment in light-elements organized in the typical 
    (anti-)correlations observed in GCs \citep[e.g.,][]{carretta09, carretta10, mucciarelli+12}. 
    \item[$\bullet$] CO and OH molecular lines are also crucial thermometers to properly estimate the star surface temperature needed for the chemical analysis. 
    \item[$\bullet$] The [C/N] and 12C/13C abundance ratios are powerful tracers of the mixing and extra-mixing processes during the red giant brach evolution.
\end{itemize}

\section{First results in Liller 1}  
\label{lil1}
\subsection{Observations and data reduction}
\label{obs}
Liller 1 is the second massive stellar system in the Galactic bulge found to host at least two distinct populations with remarkably different ages (after Terzan 5, \citealt{Ferraro_09, ferraro_16}): 12 Gyr for the oldest component, just 1-2 Gyr for the youngest one \citep{ferraro_21}. The star formation history reconstructed for this stellar system through the comparison with synthetic color-magnitude diagrams \citep{dalessandro_22} suggests a first prolonged initial burst, followed by a (low-rate) continuous star formation activity characterized by two additional events, predicting that Liller 1 stars should design a metallicity distribution with two prominent peaks, one at sub-solar and one at super-solar metallicity. Following these photometric studies, a number of spectroscopic investigations were performed to obtain a detailed chemical characterization of the hosted sub-populations. The first spectroscopic campaigns, performed at low and medium spectral resolution by using, respectively, MUSE/WFM \citep{crociati_23} and X-shooter \citep{deimer24} fully confirmed the expectation, finding a clear bimodal distribution with a sub-solar and a super-solar component. Then, \citet{fanelli+24} (hereafter F24) discussed the spectroscopic screening of 21 giants from high-resolution ($R\sim 25,000$) spectra secured at the Keck telescope with NIRSPEC, and suggested the possible presence of an additional component at intermediate metallicities. Here we present the spectroscopic analysis of 12  stars observed in Liller 1 in the context of the BulCO survey: these are the spectra at the highest spectral resolution ($R\sim 50,000$) obtained so far in this stellar system. The results provide a direct illustration of the quality of the data that we expect to secure in the next years and a taste of the science that can be performed with the BulCO survey.

The CRIRES+ data discussed in the present paper have been acquired between April 2023 and July 2024 under favorable sky conditions of clearness and seeing. Each star has been observed through the H1582 and K2166 gratings. The data reduction was performed by using the pipeline CR2RES version 1.4.1. After the standard dark and flat-field corrections, each spectrum was sky-subtracted by using nod pairs, calibrated in wavelength using arc lamps, and then extracted using the optimal extraction method, an approach that minimizes the loss of spectral resolution, maximizes the signal-to-noise ratio, and efficiently identifies local outliers or defects. The signal-to-noise ratio per resolution element of the final spectra is always $\ge$40. 
 
\begin{figure}
    \includegraphics[width=\columnwidth]{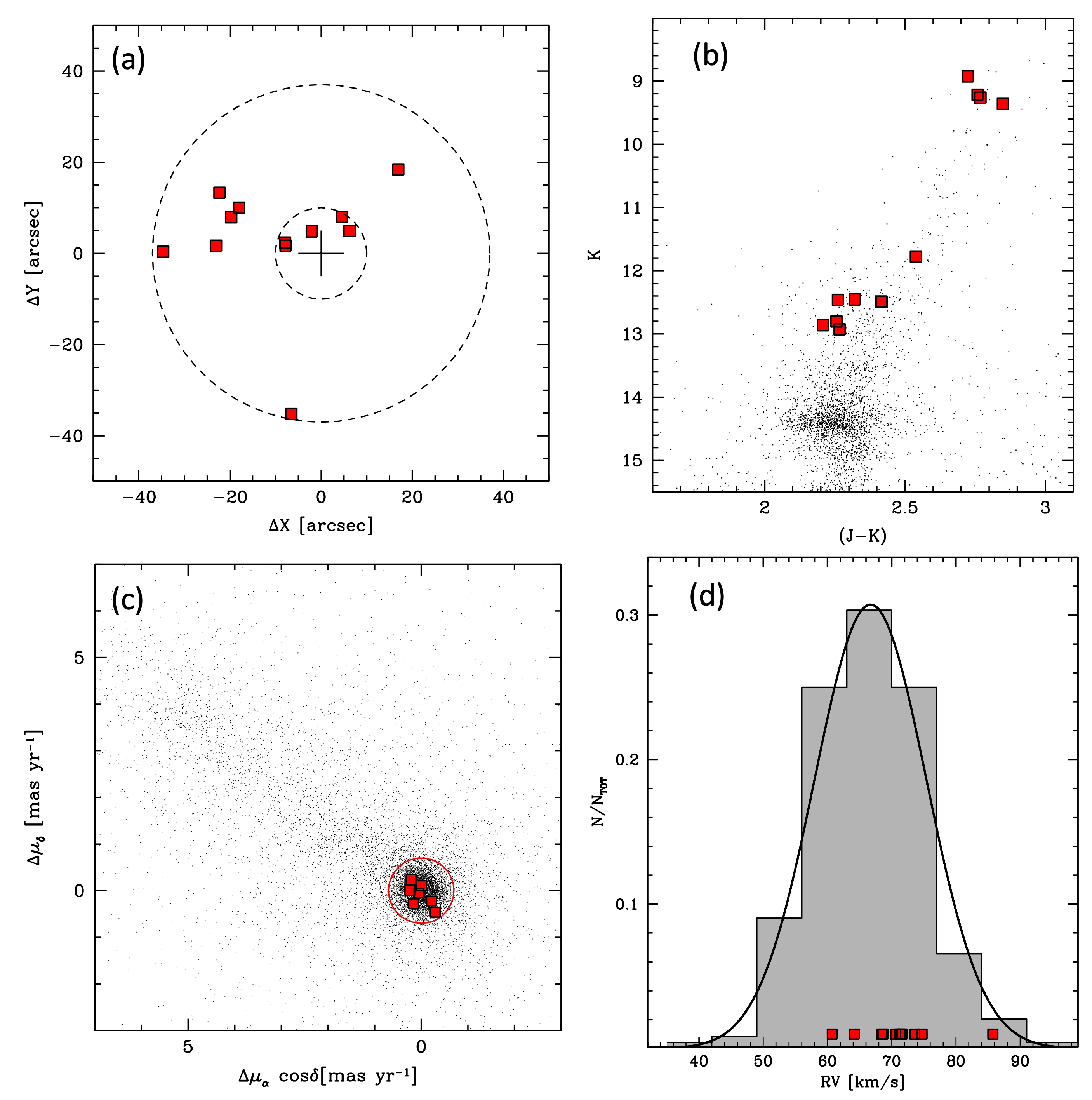}
    \caption{The spectroscopic targets (red squares) observed in  Liller 1. {\it Panel (a):} Position of the targets in the plane of the sky  with respect to the cluster center (marked with a cross). For the sake of reference, the two dashed circles mark the distances at $10\arcsec$ and $37\arcsec$ from the center.  {\it Panel (b):} Target positions in the NIR CMD.  {\it Panel (c):} Vector point diagram of relative PMs determined from HST and adaptive optics observations. The red circle delimits the stars considered as likely cluster members in \citet{dalessandro_22}. {\it Panel (d):} Radial velocity of the 12 targets compared to the overall distribution of 244 stars measured in previous studies \citep{crociati_23, deimer24, fanelli+24}.}
    \label{fig2}
\end{figure}

Table \ref{tab2} lists the coordinates, $K$-band magnitudes (from \citealt{valenti10, ferraro_21}) and physical properties of the 12 targets. Their position in the plane of the sky, with respect to the center of the system (RA=$17^{\rm h}$ $33^{\rm m}$ $24.56^{\rm s}$, Dec=$-33^\circ$ $23\arcmin$ $22.40\arcsec$; \citealp{saracino_15}) is shown in panel (a) of Figure \ref{fig2}, while their location in the 
NIR color-magnitude diagram (CMD) is plotted in panel (b). 
As can be seen, the selected stars are luminous giants, at least 1.5 magnitudes brighter than the red clump, all located in the central region of Liller 1 (at $r<37\arcsec$, corresponding to approximately one half-mass radius, $r_h=30.5\arcsec$; see \citealp{saracino_15}). To assess the membership of the observed stars via PMs, we took advantage of the detailed analysis recently presented by \citet{ferraro_21} and \citet{dalessandro_22}, where high-resolution HST and adaptive optics assisted GeMS/GEMINI images of the system, secured over a time baseline of 6.3 years, have been combined. Seven stars (namely, 100760, 300614, 300682, 300727, 400778, 400829, 400887) out of the 12 targets, have measured PMs. Their position in the vector point diagram is shown in panel (c) of Fig. \ref{fig2}. For the sake of comparison, the figure also shows the PM distribution of the surrounding stars brighter than $K=16$ (grey dots). As can be appreciated, the bulge field stars trace a prominent (almost diagonal) distribution, mainly extending at positive values of $\Delta\mu_\alpha \cos\delta$ and $\Delta\mu_\delta$ (see also Figure 2 in \citealt{dalessandro_22}), while the member stars of Liller 1 draw a well defined clump centered at (0,0). The red circle in the figure has a radius of 0.7 mas yr$^{-1}$, corresponding to 3 times the average PM error in that magnitude range.  The seven spectroscopic targets with measured PM are all located within it, as expected for likely cluster members. Unfortunately the four brightest targets (namely, 300094, 300315, 387099 and 400065) are heavily saturated in the HST images and the next brightest target (namely, 300553) lies exactly in the gaps of the GeMS detector, hence no PM can be measured for them. On the other hand no Gaia PM measurement \citep{gaia_16,gaia_23} is available for them. However, their central position  (in a region where member stars are largely dominant with respect to field interlopers; see, e.g., Figure 1 in \citealp{ferraro_21}), their position in the CMD (well aligned along the red giant branch of Liller 1), and their measured radial velocities (see next Section) guarantee that they also are high-probability cluster members.

\subsection{Atmospheric parameters and spectral analysis}
First-guess values of the surface temperature (T$_{\rm eff}$) and gravity ($\log g$) of the spectroscopic targets have been estimated photometrically from the projection of each star onto the closest isochrone in the CMD. Following \citet{ferraro_21} and \citet{dalessandro_22}, two PARSEC isochrones \citep{girardi+02,marigo+17} properly matching the old and metal-poor, and the young and metal-rich components of Liller 1 have been used: the former isochrone has an age of 12 Gyr and a metallicity [Fe/H]$=-0.3$, the latter has an age of 2 Gyr and  [Fe/H]$=+0.3$. The comparison has been performed in the differential reddening corrected CMD, adopting a distance modulus (m-M)$_0$=14.65 and an average color excess E(B-V)=4.52 \citep[see][]{pallanca_21, ferraro_21}. The first-guess values of T$_{\rm eff}$ and $\log g$ have then been spectroscopically fine-tuned by requiring the simultaneous fit of the observed OH and CO molecular lines and band-heads. The photometric and spectroscopic estimates well agree within the errors. Temperatures in the 3400- 4300 K range (with uncertainties of $\pm 100$ K), and surface gravities  in the $\log g =0.25-2.00$ interval (with a $\pm 0.3$ uncertainty) have been estimated. In addition, for all the observed stars, a microturbulence velocity of $2 \pm 0.3$ km s$^{-1}$, which is typical of giant stars of similar temperatures and metallicities, has been assumed \citep[see also][and references therein]{deimer24,fanelli+24}. 

The synthetic spectra used to measure the radial velocities and chemical abundances of the target stars have been computed by adopting the list of atomic lines from the VALD3 compilation \citep{Ryabchikova_15}, molecular lines from the website of B. Plez\footnote{\url{https://www.lupm.in2p3.fr/users/plez/}}, MARCS  atmosphere models (\citealt{gustafsson_08}), and the radiative transfer code TURBOSPECTRUM under LTE approximation \citep{alvarez_98,plez_12}. 
We generated multiple grids of synthetic spectra, with fixed stellar parameters (appropriate to each star), and varying the metallicity from $-1.0$ dex to $+0.5$ dex, in steps of 0.25 dex, with both solar-scaled and some enhancement of [$\alpha$/Fe] ($+0.3$ dex) and [N/Fe] ($+0.5$ dex) and corresponding depletion of [C/Fe] ($-0.3$ dex) for a proper computation of the molecular equilibria.
Solar-scaled [X/Fe] values have been adopted for the other elements.
The synthetic spectra computed at the nominal CRIRES+ resolution ($R\sim 50,000$) have been convoluted with a Gaussian function with a FWHM that varies between 9 and 12 km s$^{-1}$, depending on the star, to account for some additional spectral broadening due to macroturbulence, and thus optimally matching the observed spectra.

\begin{figure}
    \centering
    \includegraphics[width=\columnwidth]{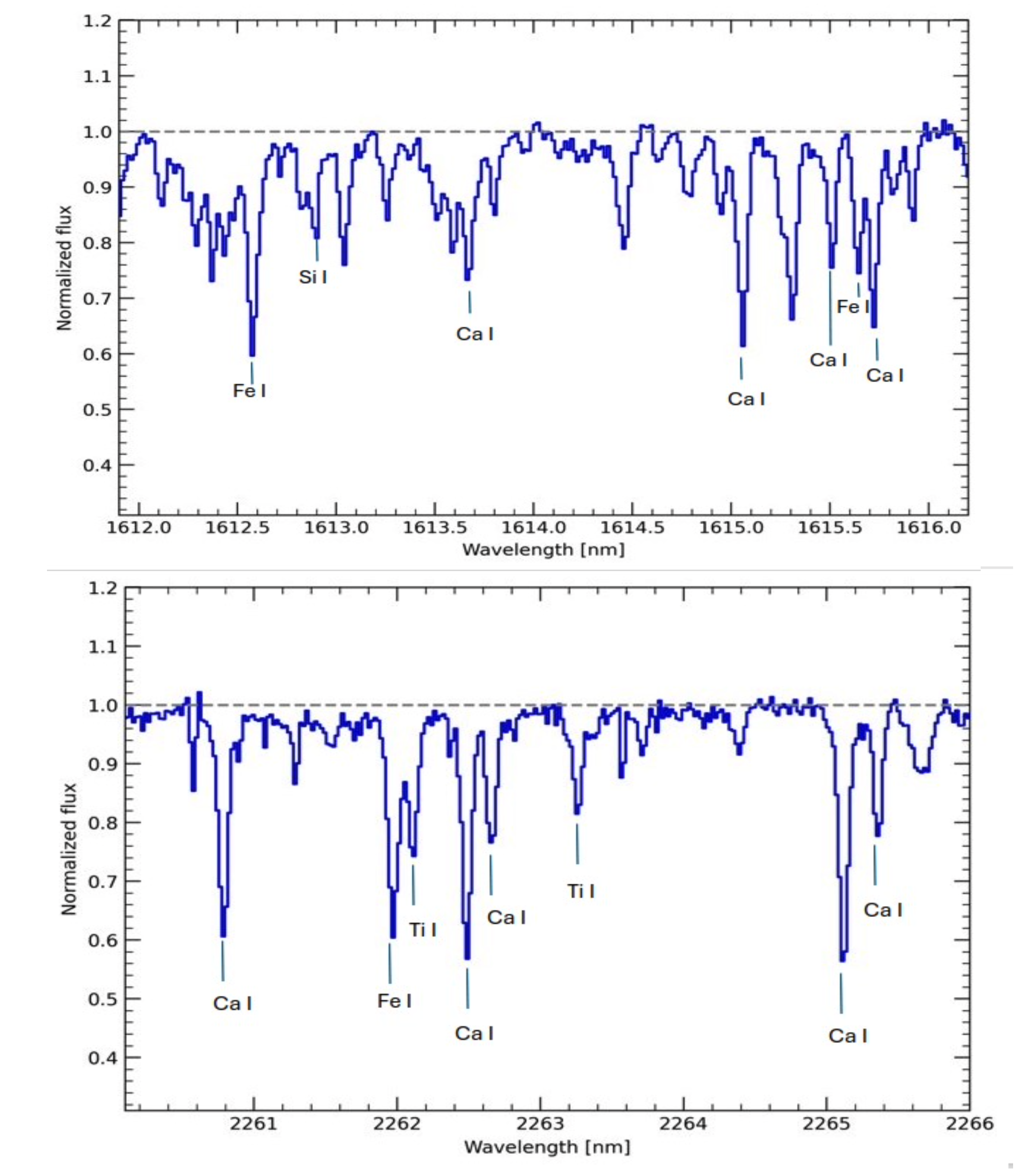}
    \caption{Region of the normalized spectra in the H (upper panel) and K (lower panel) bands obtained for star 300553. The most relevant lines for a few main elements are also marked.}
    \label{spectra}
\end{figure}

For the sake of illustration, Figure \ref{spectra} shows a small region of the normalized observed spectra in the H and K bands obtained for star 300553. The most relevant absorption lines of the main elements are also marked.

\subsection{Radial velocities and chemical abundances}
The radial velocities of the targets have been measured by means of cross-correlation between the observed and the synthetic spectra. The resulting values (see Table \ref{tab2}) range between 60 and 85 km s$^{-1}$, with a typical uncertainty of $\sim 0.5$ km s$^{-1}$. As shown in Fig. 2d, the RVs of the spectroscopic targets (red squares) well agree with the velocity distribution of the 244 stars that we already measured in Liller1 (gray histogram), which is described by a Gaussian function peaked at $66.7$ km s$^{-1}$ with dispersion $\sigma$=8.7 km s$^{-1}$ (solid line in the figure). The systemic velocity of Liller1 obtained from the entire sample we investigated so far ($66.9 \pm 0.5$ km/s) is larger than the value of 60.4$\pm$2.4 km~s$^{-1}$ quoted in \citet{baumgardt_18} based on 64 cluster members with measured radial velocities from the literature.

The chemical abundances of a first set of key elements (namely Fe, Ca, Mg, Si), determined via spectral synthesis for each selected star, are listed in Table \ref{tab2}, where only the formal errors are reported. However, the global error of each abundance ratio amounts to 0.1-0.15 dex mainly due to the uncertainties on the atmospheric parameters. Interestingly, despite the small number of stars, the sample discussed here contains all the main features of the metallicity distribution detected in Liller 1 by previous investigations (see \citealt{crociati_23, deimer24}, and F24). In fact, the majority of the stars (seven out of 12) have sub-solar metallicity, and the remaining five are super-solar. In addition, the metal-poor component shows enhanced [Ca/Fe], [Mg/Fe], and [Si/Fe] abundance ratios, while the metal-rich one displays solar-scaled values.

\section{Discussion and conclusions}
\label{conclu}
As first result of the BulCO survey currently ongoing at the ESO-VLT,  this paper presents a high-resolution spectroscopic study in the H and K bands of 12 giant stars observed in the bulge complex system Liller 1. 
To best illustrate the results in the context of the current knowledge of the chemical properties of the system, in the following we discuss the entire sample of stars for which the abundances of Fe, Ca, Mg, and Si have been measured from high- and mid-resolution spectra. As mentioned above, F24 presented the chemical abundances of 21 giants observed at $R\sim 25,000$ with NIRSPEC at keck, while \citet{deimer24} determined the abundances of almost the same chemical elements in 27 giants observed at $R\sim 8,000$ with XSHOOTER at ESO-VLT, nine of which are in common with F24. Hence, we have in hand a total sample of 51 stars, observed at high and intermediate resolution, for which the abundances of these elements are available. In addition, \citet{crociati_23} measured the iron abundance of 53 stars from the Calcium triplet lines detected in MUSE spectra ($R\sim 3100$).

\begin{figure}
    \centering
    \includegraphics[width=\columnwidth]{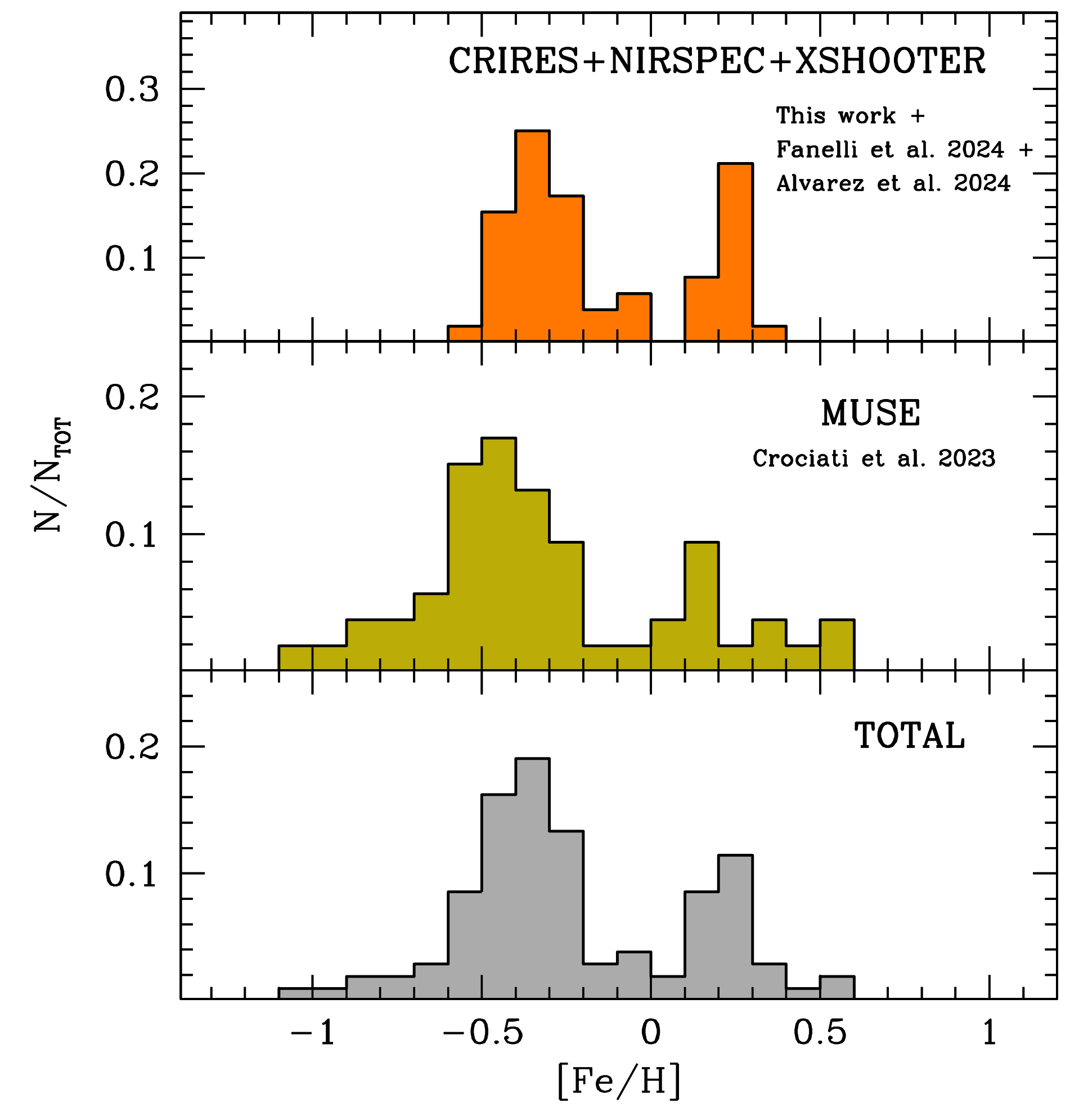}
    \caption{Metallicity distribution of the stars of Liller 1 obtained from recent surveys. {\it Top panel}: the results of this work (from CRIRES+ at VLT), F24 (from NIRSPEC at Keck) and those of \citet[][from XSHOOTER at VLT]{deimer24}. {\it Central panel}: the iron distribution obtained from MUSE at VLT observations \citealt{crociati_23}. {\it Bottom panel}: The [Fe/H] distribution of the total sample obtained by combining the different datasets and counting 105 stars. }
    \label{hist}
\end{figure}

Figure \ref{hist} summarizes the current knowledge about the iron distribution in Liller 1. 
The results of the present analysis have been merged with those of F24 and \citet{deimer24} in the top panel, since these are the spectra with the highest spectral resolution. All the datasets clearly show a well defined bimodal distribution, and the total sample is characterized by a dominant sub-solar component peaking at [Fe/H]=$-0.36\pm 0.03$ and with a 1$\sigma$ dispersion of 0.15$\pm$0.02, counting approximately 71\% of the measured stars, and a super-solar one counting the remaining 29\% of the population, with a peak at [Fe/H]=$+0.24 \pm 0.02$ and $1 \sigma =0.10 \pm 0.02$. Thus, the presence of multiple stellar components with different iron abundances that was predicted on the basis of the photometric properties of Liller 1 \citep{ferraro_21,dalessandro_22} is now solidly confirmed by multi-instrument spectroscopic campaigns. This result is in good agreement with the expectations from the star formation history of the system determined in the context of a self-enrichment scenario \citep{dalessandro_22}, where $\sim 70$\% of the total mass of the system is formed in the first major burst, and the remaining 30\% is assembled over the subsequent stellar system lifetime.

\begin{figure*}
    \centering
    \includegraphics[width=0.99\linewidth]{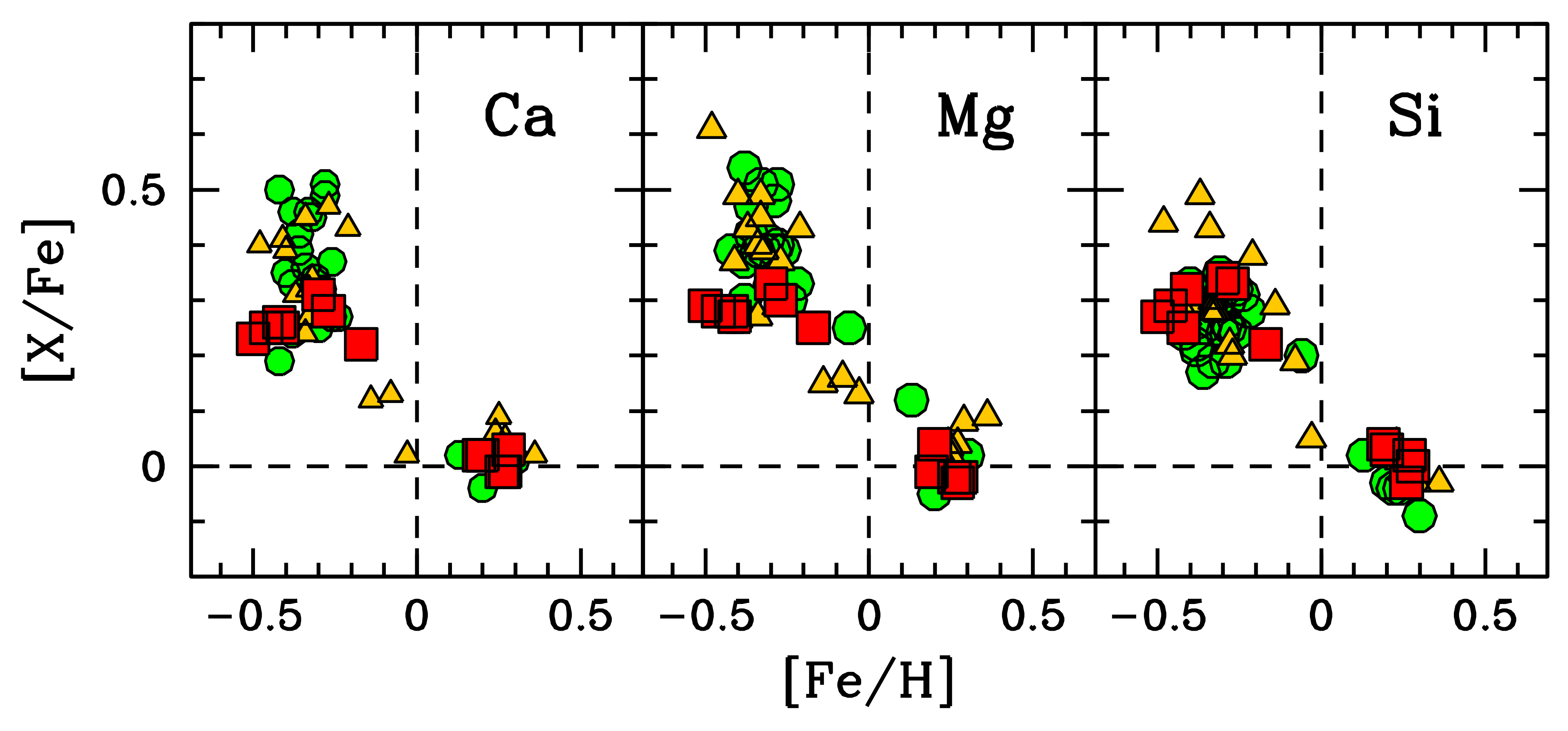}
    \caption{[Ca/Fe], [Mg/Fe] and [Si/Fe] as a function of [Fe/H] for the 51 stars of Liller 1 measured so far.  The values obtained from different surveys at high- and mid- spectral resolution are shown with different symbols: the red squares are the 12 stars observed with CRIRES+ and presented in this paper, the yellow triangles are from the NIRSPEC sample (F24), and the green circles are from the XSHOOTER sample \citep{deimer24}. For reference, the dashed vertical and horizontal lines mark the solar values.}
    \label{ratio}
\end{figure*}

The chemical patterns measured in the [$\alpha$/Fe]$-$[Fe/H] diagram allow us to go deeper into the characterization of these multi-iron sub-populations, and obtain key indications about their possible formation scenarios. In fact,  the abundance patterns in this diagram are so distinctive that they can be used as ``DNA test" of the stellar population origin. This is due to the fact that $\alpha$-elements (like calcium, magnesium, silicon, etc.) are mainly synthesized by massive stars and released to the interstellar medium over a short time scale ($10^7$ yr) by the explosion of core collapse SNeII. While a small amount of iron is also produced by SNeII, the large majority is instead ejected by SNIa thermonuclear explosions, occurring over longer timescales \citep[e.g.,][]{matteucci_recchi2001}. This later injection of iron implies that the [$\alpha$/Fe] ratio, initially set and kept nearly constant by SNeII, starts to rapidly decrease when the first generation of SNeIa explodes, thus producing a characteristic knee in the [$\alpha$/Fe]-[Fe/H] diagram. The knee therefore flags the value of [Fe/H] that the  interstellar medium achieved (due to the action of SNeII only) at the epoch of the first SNIa explosions. This implies that the [$\alpha$/Fe]$-$[Fe/H] pattern is a powerful indicator of the SFR in the environment where stars formed: the higher is the SFR, the larger is the contribution of SNeII to the chemical enrichment before the bulk of SNeIa explode, thus the larger is the metallicity of the knee in the diagram. The [$\alpha$/Fe]$-$[Fe/H] diagram obtained for the total sample is plotted in Figure \ref{ratio}. All the available datasets clearly show that the stellar components with different iron content also have different levels of [$\alpha$/Fe] enhancement, with the most metal-poor sub-population being enhanced (on average, by a factor of 2-3) in [Ca/Fe], [Mg/Fe], and [Si/Fe], and the super-solar one having about solar-scaled values. This testifies that they formed from differently enriched  interstellar medium: the main component, with sub-solar iron and [$\alpha$/Fe] enhancement, formed (likely at an early epoch) from gas enriched by SNeII, while the one with super-solar iron and solar-scaled [$\alpha$/Fe] has been generated (likely at much later epochs) from gas also enriched by SNeIa on a longer timescale. Interestingly, a few stars have been also detected in-between the main metal-poor and metal-rich sub-groups, displaying intermediate iron content ($-0.2<$[Fe/H]$<+0.1$ dex) and also intermediate enhancement in [Ca/Fe], [Mg/Fe], and [Si/Fe]. This is another very important feature providing insights into the formation history of Liller 1. In fact, the star formation history reconstructed from the analysis of the CMD \citep{dalessandro_22} predicts that Liller 1 has been active in forming stars for its entire lifetime, with three main episodes: a first, broad event occurred 12-13 Gyr ago, followed by a long tail of star formation activity at lower rate, with an intermediate (6-8 Gyr ago) and a very recent (2 Gyr ago) bursts. Of course, this would have favored the formation of a broad distribution in metallicity and $\alpha$-elements, keeping memory not only of the main star formation episodes, but also of the overall (low-rate) activity, 
which is now observable in the presence of a few stars with intermediate values of [Fe/H] and [$\alpha$/Fe].

\begin{figure*}
    \centering
    \includegraphics[width=0.98\linewidth]{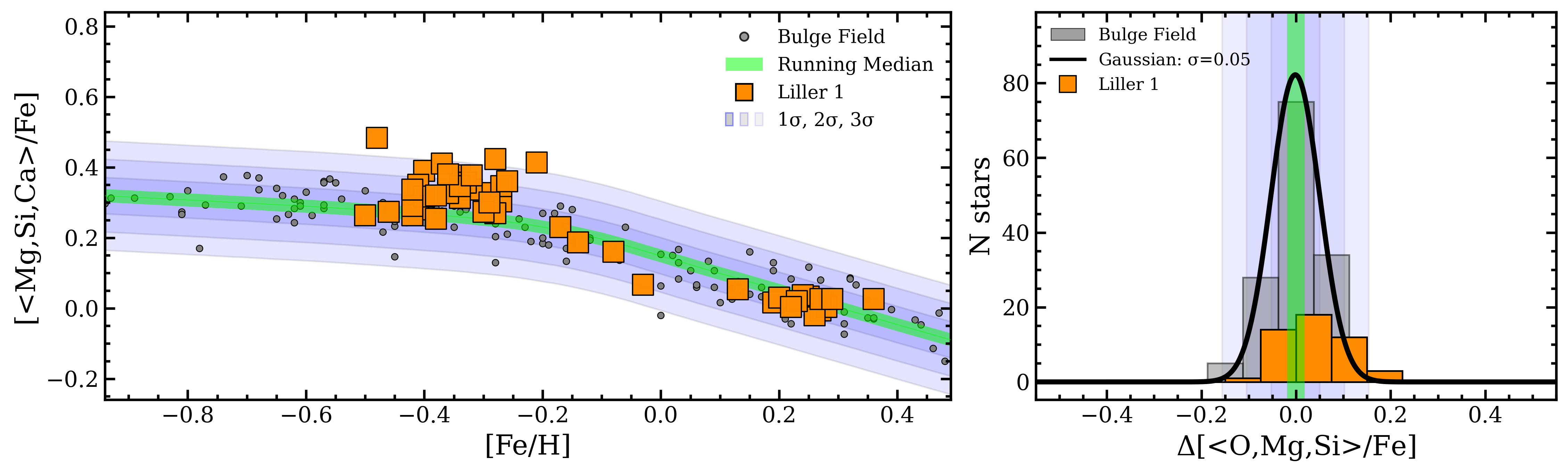}
    \caption{Behavior of the average $\alpha-$element abundance ratio [<Ca,Mg,O>/Fe] as a function of the metallicity ([Fe/H]) for the total of 51 stars observed so far in Liller 1 (large orange squares), compared with the distribution observed for bulge field stars (small gray circles, from \citealt{johnson_14}). The green line traces the LOWESS median line of the bulge star distribution, and the 1, 2, and 3~$\sigma$ regions are also plotted with different color shading. The right panel shows the distribution of the distances from the LOWESS median line measured for the bulge field stars (gray histogram) and the Liller 1 stars (orange histogram). The Gaussian function reproducing the bulge distribution is also shown as a black curve.}
    \label{alpha-fe}
\end{figure*}
With the data in hand, we push further our investigation and make a preliminary comparison of the chemical DNA of Liller1 with that of the Galactic bulge. For each target shown in Fig. \ref{ratio}, we determined the average abundance of the three analyzed $\alpha$-elements (Ca, Mg and Si), and compared its dependence on [Fe/H] with the pattern obtained from high-resolution optical and NIR spectroscopy of about 150 bulge field stars  \citep{johnson_14}. 
The latter abundances have been placed on the \citet{magg_22} solar reference. The result is plotted in Figure \ref{alpha-fe} (left panel) and reveals an astonishing similarity: the two distributions follow exactly the same pattern. This is fully confirmed by a simple statistical analysis. According to the methodology applied, e.g., in Origlia et al. (2025), we modeled the bulge distribution using a Locally Weighted Scatterplot Smoothing\footnote{\url{https://www.statsmodels.org/devel/generated/statsmodels.nonparametric. smoothers lowess.lowess.html}} (LOWESS) function (green line in the left panel of Fig. \ref{alpha-fe}), which is a non-parametric regression technique to obtain an optimal median curve. The right panel of the same figure plots the distribution of the distances of the bulge stars (grey histogram) and Liller 1 stars (orange histogram) from the LOWESS curve. As can be appreciated, the vast majority of the Liller1 stars is included within $3\sigma $ from the bulge distributions, with 77\% of them lying within 2$\sigma$, and 49\% located within 1$\sigma$ only. This clearly assesses the full compatibility of the two samples. As it is well known, the abundance pattern defined by Galactic bulge stars in the [$\alpha$/Fe]$-$[Fe/H] diagram is unique in the Local Universe, since it is characterized by a knee at metallicity [Fe/H]$\sim -0.3$ that is significantly higher than those observed in the MW halo and disk ([Fe/H]$\sim -1.0$), and in local dwarf galaxies ([Fe/H]$\sim  -1.5$; see, e.g., Figure 4 in \citealt{matteucci90}, Figure 11 in \citealt{tolstoy09}, and Figure 6 in \citealp{kobayashi23}). This clearly testifies an environment that experienced a particularly intense SFR and, because of its uniqueness in the Local Universe, this pattern can be considered as the chemical DNA of the MW bulge. The results shown in Fig. \ref{alpha-fe} therefore unequivocally demonstrates the kinship of Liller 1 with the central region of our galaxy.

Although this analysis is performed on a limited number of stars (51, which however counts the entire sample of abundances obtained so far from high/mid-resolution spectroscopy), it already allows us to shed new light into the formation process of Liller 1. Indeed, the evidence that Liller 1 shares the same chemical DNA of the Galactic bulge solidly locates the birth place of this stellar system into the MW core, and definitely excludes any scenario invoking an accretion of a satellite from outside the bulge. This is also supported by the reconstructed orbit (as computed from Gaia proper motions and line-of-sight velocities by \citealt{baumgardt_18}),\footnote{Fundamental parameters of Galactic globular clusters, \url{https://people.smp.uq.edu.au/HolgerBaumgardt/globular/}} which confirms that Liller 1 spent all its life confined within the bulge. In addition, the increasing evidence of a continuous (instead of double-peaked) distribution of Liller 1 stars in the [$\alpha$/Fe]$-$[Fe/H] diagram challenges the scenarios invoking the merging between two GCs\footnote{Indeed, this scenario is also strongly disfavored by the fact that not a single GC as metal-rich as the super-solar population of Liller 1 is known in our galaxy, since all Galactic GCs have [Fe/H]$<0$ \citep[see, e.g.,][]{harris96}.} \citep{khoperskov18, mastrobuono19, pfeffer21}, or the accretion of a giant molecular cloud by a genuine globular \citep{mckenzie18, bastian_22}, since these events should happen at most once in a cluster lifetime. Instead, the difference in the $\alpha-$enhancement observed for the multi-iron sub-populations is suggestive of a self-enrichment scenario, which is also  supported by the fact that the most metal-rich component is more centrally concentrated than the most metal-poor one \citep{ferraro_21, dalessandro_22, crociati_23}. Within the variety of fossil relics expected to populate the bulge, the photometric and spectroscopic pieces of evidence collected so far all seem to concur in suggesting that Liller 1 is the remnant of a much more massive stellar system that formed in-situ at the epoch of the Galaxy assembling and shared a common evolutionary history with the bulge itself (a possible BFF).

Intriguingly, also in the case of Terzan 5, all the photometric, kinematic and chemical properties observed so far have provided robust evidence of a bulge in-situ formation and evolution (see \citealt{origlia+25} and references therein). Thus, two possible BFFs have been discovered so far in the Galactic bulge.
Thanks to their initially large mass, these primordial systems were able to retain the SN ejecta and experience an almost continuous, low-rate star formation, with a few more intense bursts from which iron-enriched stars with different $\alpha-$enhancement had origin. A chemical enriching model (similar to the one discussed for Terzan 5 in \citealt{romano23}) specifically designed for Liller 1 is foreseen, with the goal to determine the properties that the progenitor of this system should have had in order to originate its currently observed characteristics. Of course, a more detailed chemical screening of the stars in Liller 1 is of paramount importance to shed light on these topics. This is one of the aims of the ongoing BulCO survey, which is expected to acquire $R=50,000$ resolution spectra for a total of more than 50 giant stars in Liller 1 in the next months. This sample will provide the definitive shape of the distribution in the [$\alpha$/Fe]$-$[Fe/H] diagram, also characterizing the currently under-populated region around solar metallicity. Furthermore, chemical abundances for elements originating from different nucleosynthetic paths will be available, thus removing some important degeneracies plaguing chemical evolution models. We will thus be in the optimal conditions to unravel the formation history of Liller 1 in the context of the Milky Way bulge assembly.

\begin{acknowledgements}
This work is part of the project Cosmic-Lab at the Physics and Astronomy Department “A. Righi” of the Bologna University (\url{http:// www.cosmic-lab.eu/ Cosmic-Lab/Home.html}). CF and LO acknowledge the financial support by INAF within the VLT-MOONS project. A.M.  and D.R. acknowledge support from the project "LEGO – Reconstructing the building blocks of the Galaxy by chemical tagging" (P.I. A. Mucciarelli) granted by the Italian MUR through contract PRIN 2022LLP8TK\_001.
\end{acknowledgements}

\bibliographystyle{aa} % style aa.bst
\bibliography{biblio} % your references Yourfile.bib

\begin{thebibliography}{102}
\expandafter\ifx\csname natexlab\endcsname\relax\def\natexlab#1{#1}\fi

\bibitem[{{Alvarez} \& {Plez}(1998)}]{alvarez_98}
{Alvarez}, R. \& {Plez}, B. 1998, \aap, 330, 1109

\bibitem[{{Alvarez Garay} {et~al.}(2024{\natexlab{a}}){Alvarez Garay},
  {Fanelli}, {Origlia}, {Pallanca}, {Mucciarelli}, {Chiappino}, {Crociati},
  {Lanzoni}, {Ferraro}, {Rich}, \& {Dalessandro}}]{deimer24}
{Alvarez Garay}, D.~A., {Fanelli}, C., {Origlia}, L., {et~al.}
  2024{\natexlab{a}}, \aap, 686, A198

\bibitem[{{Alvarez Garay} {et~al.}(2024{\natexlab{b}}){Alvarez Garay},
  {Mucciarelli}, {Bellazzini}, {Lardo}, \& {Ventura}}]{deimer2024_omega}
{Alvarez Garay}, D.~A., {Mucciarelli}, A., {Bellazzini}, M., {Lardo}, C., \&
  {Ventura}, P. 2024{\natexlab{b}}, \aap, 681, A54

\bibitem[{{Alvarez Garay} {et~al.}(2022){Alvarez Garay}, {Mucciarelli},
  {Lardo}, {Bellazzini}, \& {Merle}}]{deimer2022_omega}
{Alvarez Garay}, D.~A., {Mucciarelli}, A., {Lardo}, C., {Bellazzini}, M., \&
  {Merle}, T. 2022, \apjl, 928, L11

\bibitem[{{Bastian} \& {Pfeffer}(2022)}]{bastian_22}
{Bastian}, N. \& {Pfeffer}, J. 2022, \mnras, 509, 614

\bibitem[{{Baumgardt} \& {Hilker}(2018)}]{baumgardt_18}
{Baumgardt}, H. \& {Hilker}, M. 2018, \mnras, 478, 1520

\bibitem[{{Bekki} \& {Freeman}(2003)}]{bekki+03}
{Bekki}, K. \& {Freeman}, K.~C. 2003, \mnras, 346, L11

\bibitem[{{Bellini} {et~al.}(2017){Bellini}, {Milone}, {Anderson}, {Marino},
  {Piotto}, {van der Marel}, {Bedin}, \& {King}}]{bellini+17}
{Bellini}, A., {Milone}, A.~P., {Anderson}, J., {et~al.} 2017, \apj, 844, 164

\bibitem[{{Bournaud}(2016)}]{bournaud16}
{Bournaud}, F. 2016, in Astrophysics and Space Science Library, Vol. 418,
  Galactic Bulges, ed. E.~{Laurikainen}, R.~{Peletier}, \& D.~{Gadotti}, 355

\bibitem[{{Bournaud} \& {Elmegreen}(2009)}]{bournaud09}
{Bournaud}, F. \& {Elmegreen}, B.~G. 2009, \apjl, 694, L158

\bibitem[{{Cadelano} {et~al.}(2023){Cadelano}, {Pallanca}, {Dalessandro},
  {Salaris}, {Mucciarelli}, {Leanza}, {Ferraro}, {Lanzoni}, {Rosie Chen},
  {Freire}, {Heinke}, \& {Ransom}}]{cadelano23}
{Cadelano}, M., {Pallanca}, C., {Dalessandro}, E., {et~al.} 2023, \aap, 679,
  L13

\bibitem[{{Callingham} {et~al.}(2022){Callingham}, {Cautun}, {Deason}, {Frenk},
  {Grand}, \& {Marinacci}}]{callingham+22}
{Callingham}, T.~M., {Cautun}, M., {Deason}, A.~J., {et~al.} 2022, \mnras, 513,
  4107

\bibitem[{{Carretta} \& {Bragaglia}(2023)}]{carretta23}
{Carretta}, E. \& {Bragaglia}, A. 2023, \aap, 677, A73

\bibitem[{{Carretta} {et~al.}(2009){Carretta}, {Bragaglia}, {Gratton},
  {Lucatello}, {Catanzaro}, {Leone}, {Bellazzini}, {Claudi}, {D'Orazi},
  {Momany}, {Ortolani}, {Pancino}, {Piotto}, {Recio-Blanco}, \&
  {Sabbi}}]{carretta09}
{Carretta}, E., {Bragaglia}, A., {Gratton}, R.~G., {et~al.} 2009, \aap, 505,
  117

\bibitem[{{Carretta} {et~al.}(2010){Carretta}, {Bragaglia}, {Gratton},
  {Recio-Blanco}, {Lucatello}, {D'Orazi}, \& {Cassisi}}]{carretta10}
{Carretta}, E., {Bragaglia}, A., {Gratton}, R.~G., {et~al.} 2010, \aap, 516,
  A55

\bibitem[{{Ceccarelli} {et~al.}(2024){Ceccarelli}, {Mucciarelli}, {Massari},
  {Bellazzini}, \& {Matsuno}}]{ceccarelli+24}
{Ceccarelli}, E., {Mucciarelli}, A., {Massari}, D., {Bellazzini}, M., \&
  {Matsuno}, T. 2024, \aap, 691, A226

\bibitem[{{Crociati} {et~al.}(2024){Crociati}, {Cignoni}, {Dalessandro},
  {Pallanca}, {Massari}, {Ferraro}, {Lanzoni}, {Origlia}, \&
  {Valenti}}]{crociati+24}
{Crociati}, C., {Cignoni}, M., {Dalessandro}, E., {et~al.} 2024, \aap, 691,
  A311

\bibitem[{{Crociati} {et~al.}(2023){Crociati}, {Valenti}, {Ferraro},
  {Pallanca}, {Lanzoni}, {Cadelano}, {Fanelli}, {Origlia}, {Leanza},
  {Dalessandro}, {Mucciarelli}, \& {Rich}}]{crociati_23}
{Crociati}, C., {Valenti}, E., {Ferraro}, F.~R., {et~al.} 2023, \apj, 951, 17

\bibitem[{{Dalessandro} {et~al.}(2022){Dalessandro}, {Crociati}, {Cignoni},
  {Ferraro}, {Lanzoni}, {Origlia}, {Pallanca}, {Rich}, {Saracino}, \&
  {Valenti}}]{dalessandro_22}
{Dalessandro}, E., {Crociati}, C., {Cignoni}, M., {et~al.} 2022, \apj, 940, 170

\bibitem[{{Dalessandro} {et~al.}(2016){Dalessandro}, {Saracino}, {Origlia},
  {Marchetti}, {Ferraro}, {Lanzoni}, {Geisler}, {Cohen}, {Mauro}, \&
  {Villanova}}]{dalessandro16}
{Dalessandro}, E., {Saracino}, S., {Origlia}, L., {et~al.} 2016, \apj, 833, 111

\bibitem[{{Davis} {et~al.}(1985){Davis}, {Efstathiou}, {Frenk}, \&
  {White}}]{davis85}
{Davis}, M., {Efstathiou}, G., {Frenk}, C.~S., \& {White}, S.~D.~M. 1985, \apj,
  292, 371

\bibitem[{{Deras} {et~al.}(2023){Deras}, {Cadelano}, {Ferraro}, {Lanzoni}, \&
  {Pallanca}}]{deras23}
{Deras}, D., {Cadelano}, M., {Ferraro}, F.~R., {Lanzoni}, B., \& {Pallanca}, C.
  2023, \apj, 942, 104

\bibitem[{{Dorn} {et~al.}(2014){Dorn}, {Anglada-Escude}, {Baade}, {Bristow},
  {Follert}, {Gojak}, {Grunhut}, {Hatzes}, {Heiter}, {Hilker}, {Ives}, {Jung},
  {K{\"a}ufl}, {Kerber}, {Klein}, {Lizon}, {Lockhart}, {L{\"o}winger},
  {Marquart}, {Oliva}, {Origlia}, {Pasquini}, {Paufique}, {Piskunov}, {Pozna},
  {Reiners}, {Smette}, {Smoker}, {Seemann}, {Stempels}, \& {Valenti}}]{dorn+14}
{Dorn}, R.~J., {Anglada-Escude}, G., {Baade}, D., {et~al.} 2014, The Messenger,
  156, 7

\bibitem[{{Dorn} {et~al.}(2023){Dorn}, {Bristow}, {Smoker}, {Rodler}, {Lavail},
  {Accardo}, {van den Ancker}, {Baade}, {Baruffolo}, {Courtney-Barrer},
  {Blanco}, {Brucalassi}, {Cumani}, {Follert}, {Haimerl}, {Hatzes}, {Haug},
  {Heiter}, {Hinterschuster}, {Hubin}, {Ives}, {Jung}, {Jones}, {Kaeufl},
  {Kirchbauer}, {Klein}, {Kochukhov}, {Korhonen}, {K{\"o}hler}, {Lizon},
  {Moins}, {Molina-Conde}, {Marquart}, {Neeser}, {Oliva}, {Pallanca},
  {Pasquini}, {Paufique}, {Piskunov}, {Reiners}, {Schneller}, {Schmutzer},
  {Seemann}, {Slumstrup}, {Smette}, {Stegmeier}, {Stempels}, {Tordo},
  {Valenti}, {Valenzuela}, {Vernet}, {Vinther}, \& {Wehrhahn}}]{dorn+23}
{Dorn}, R.~J., {Bristow}, P., {Smoker}, J.~V., {et~al.} 2023, \aap, 671, A24

\bibitem[{{Elmegreen} {et~al.}(2008){Elmegreen}, {Bournaud}, \&
  {Elmegreen}}]{elme08}
{Elmegreen}, B.~G., {Bournaud}, F., \& {Elmegreen}, D.~M. 2008, \apj, 688, 67

\bibitem[{{Elmegreen} {et~al.}(2009){Elmegreen}, {Elmegreen}, {Fernandez}, \&
  {Lemonias}}]{elme09}
{Elmegreen}, B.~G., {Elmegreen}, D.~M., {Fernandez}, M.~X., \& {Lemonias},
  J.~J. 2009, \apj, 692, 12

\bibitem[{{Fanelli} {et~al.}(2024){Fanelli}, {Origlia}, {Rich}, {Ferraro},
  {Alvarez Garay}, {Chiappino}, {Lanzoni}, {Pallanca}, {Crociati}, \&
  {Dalessandro}}]{fanelli+24}
{Fanelli}, C., {Origlia}, L., {Rich}, R.~M., {et~al.} 2024, \aap, 690, A139

\bibitem[{{Ferraro} {et~al.}(2009){Ferraro}, {Dalessandro}, {Mucciarelli},
  {Beccari}, {Rich}, {Origlia}, {Lanzoni}, {Rood}, {Valenti}, {Bellazzini},
  {Ransom}, \& {Cocozza}}]{Ferraro_09}
{Ferraro}, F.~R., {Dalessandro}, E., {Mucciarelli}, A., {et~al.} 2009, \nat,
  462, 483

\bibitem[{{Ferraro} {et~al.}(2016){Ferraro}, {Massari}, {Dalessandro},
  {Lanzoni}, {Origlia}, {Rich}, \& {Mucciarelli}}]{ferraro_16}
{Ferraro}, F.~R., {Massari}, D., {Dalessandro}, E., {et~al.} 2016, \apj, 828,
  75

\bibitem[{{Ferraro} {et~al.}(2000){Ferraro}, {Montegriffo}, {Origlia}, \& {Fusi
  Pecci}}]{ferraro00}
{Ferraro}, F.~R., {Montegriffo}, P., {Origlia}, L., \& {Fusi Pecci}, F. 2000,
  \aj, 119, 1282

\bibitem[{{Ferraro} {et~al.}(2018){Ferraro}, {Mucciarelli}, {Lanzoni},
  {Pallanca}, {Lapenna}, {Origlia}, {Dalessandro}, {Valenti}, {Beccari},
  {Bellazzini}, {Vesperini}, {Varri}, \& {Sollima}}]{ferraro18}
{Ferraro}, F.~R., {Mucciarelli}, A., {Lanzoni}, B., {et~al.} 2018, \apj, 860,
  50

\bibitem[{{Ferraro} {et~al.}(2021){Ferraro}, {Pallanca}, {Lanzoni}, {Crociati},
  {Dalessandro}, {Origlia}, {Rich}, {Saracino}, {Mucciarelli}, {Valenti},
  {Geisler}, {Mauro}, {Villanova}, {Moni Bidin}, \& {Beccari}}]{ferraro_21}
{Ferraro}, F.~R., {Pallanca}, C., {Lanzoni}, B., {et~al.} 2021, Nature
  Astronomy, 5, 311

\bibitem[{{Ferraro} {et~al.}(2004){Ferraro}, {Sollima}, {Pancino},
  {Bellazzini}, {Straniero}, {Origlia}, \& {Cool}}]{ferraro+04}
{Ferraro}, F.~R., {Sollima}, A., {Pancino}, E., {et~al.} 2004, \apjl, 603, L81

\bibitem[{{Gaia Collaboration} {et~al.}(2016){Gaia Collaboration}, {Prusti},
  {de Bruijne}, {Brown}, {Vallenari}, {Babusiaux}, {et~al.}}]{gaia_16}
{Gaia Collaboration}, {Prusti}, T., {de Bruijne}, J.~H.~J., {et~al.} 2016,
  \aap, 595, A1

\bibitem[{{Gaia Collaboration} {et~al.}(2023){Gaia Collaboration}, {Vallenari},
  {Brown}, {Prusti}, {de Bruijne}, {Arenou}, {et~al.}}]{gaia_23}
{Gaia Collaboration}, {Vallenari}, A., {Brown}, A.~G.~A., {et~al.} 2023, \aap,
  674, A1

\bibitem[{{Genzel} {et~al.}(2011){Genzel}, {Newman}, {Jones}, {F{\"o}rster
  Schreiber}, {Shapiro}, {Genel}, {Lilly}, {Renzini}, {Tacconi}, {Bouch{\'e}},
  {Burkert}, {Cresci}, {Buschkamp}, {Carollo}, {Ceverino}, {Davies}, {Dekel},
  {Eisenhauer}, {Hicks}, {Kurk}, {Lutz}, {Mancini}, {Naab}, {Peng},
  {Sternberg}, {Vergani}, \& {Zamorani}}]{genzel11}
{Genzel}, R., {Newman}, S., {Jones}, T., {et~al.} 2011, \apj, 733, 101

\bibitem[{{Girardi} {et~al.}(2002){Girardi}, {Bertelli}, {Bressan}, {Chiosi},
  {Groenewegen}, {Marigo}, {Salasnich}, \& {Weiss}}]{girardi+02}
{Girardi}, L., {Bertelli}, G., {Bressan}, A., {et~al.} 2002, \aap, 391, 195

\bibitem[{{Gustafsson} {et~al.}(2008){Gustafsson}, {Edvardsson}, {Eriksson},
  {J{\o}rgensen}, {Nordlund}, \& {Plez}}]{gustafsson_08}
{Gustafsson}, B., {Edvardsson}, B., {Eriksson}, K., {et~al.} 2008, \aap, 486,
  951

\bibitem[{{Harris}(1996)}]{harris96}
{Harris}, W.~E. 1996, \aj, 112, 1487

\bibitem[{{Helmi}(2020)}]{helmi20}
{Helmi}, A. 2020, \araa, 58, 205

\bibitem[{{Immeli} {et~al.}(2004){Immeli}, {Samland}, {Gerhard}, \&
  {Westera}}]{immeli_04}
{Immeli}, A., {Samland}, M., {Gerhard}, O., \& {Westera}, P. 2004, \aap, 413,
  547

\bibitem[{{Je{\v{r}}{\'a}bkov{\'a}} {et~al.}(2018){Je{\v{r}}{\'a}bkov{\'a}},
  {Hasani Zonoozi}, {Kroupa}, {Beccari}, {Yan}, {Vazdekis}, \&
  {Zhang}}]{jera18}
{Je{\v{r}}{\'a}bkov{\'a}}, T., {Hasani Zonoozi}, A., {Kroupa}, P., {et~al.}
  2018, \aap, 620, A39

\bibitem[{{Johnson} \& {Pilachowski}(2010)}]{johnson2010_omega}
{Johnson}, C.~I. \& {Pilachowski}, C.~A. 2010, \apj, 722, 1373

\bibitem[{{Johnson} {et~al.}(2014){Johnson}, {Rich}, {Kobayashi}, {Kunder}, \&
  {Koch}}]{johnson_14}
{Johnson}, C.~I., {Rich}, R.~M., {Kobayashi}, C., {Kunder}, A., \& {Koch}, A.
  2014, \aj, 148, 67

\bibitem[{{Kaeufl} {et~al.}(2004){Kaeufl}, {Ballester}, {Biereichel},
  {Delabre}, {Donaldson}, {Dorn}, {Fedrigo}, {Finger}, {Fischer}, {Franza},
  {Gojak}, {Huster}, {Jung}, {Lizon}, {Mehrgan}, {Meyer}, {Moorwood}, {Pirard},
  {Paufique}, {Pozna}, {Siebenmorgen}, {Silber}, {Stegmeier}, \&
  {Wegerer}}]{kaufl+04}
{Kaeufl}, H.-U., {Ballester}, P., {Biereichel}, P., {et~al.} 2004, in Society
  of Photo-Optical Instrumentation Engineers (SPIE) Conference Series, Vol.
  5492, Ground-based Instrumentation for Astronomy, ed. A.~F.~M. {Moorwood} \&
  M.~{Iye}, 1218--1227

\bibitem[{{Khoperskov} {et~al.}(2018){Khoperskov}, {Mastrobuono-Battisti}, {Di
  Matteo}, \& {Haywood}}]{khoperskov18}
{Khoperskov}, S., {Mastrobuono-Battisti}, A., {Di Matteo}, P., \& {Haywood}, M.
  2018, \aap, 620, A154

\bibitem[{{Kobayashi} {et~al.}(2020){Kobayashi}, {Karakas}, \&
  {Lugaro}}]{koba20}
{Kobayashi}, C., {Karakas}, A.~I., \& {Lugaro}, M. 2020, \apj, 900, 179

\bibitem[{{Kobayashi} \& {Taylor}(2023)}]{kobayashi23}
{Kobayashi}, C. \& {Taylor}, P. 2023, arXiv e-prints, arXiv:2302.07255

\bibitem[{{Lanzoni} {et~al.}(2007){Lanzoni}, {Dalessandro}, {Ferraro},
  {Miocchi}, {Valenti}, \& {Rood}}]{lanzoni07}
{Lanzoni}, B., {Dalessandro}, E., {Ferraro}, F.~R., {et~al.} 2007, \apjl, 668,
  L139

\bibitem[{{Lanzoni} {et~al.}(2010){Lanzoni}, {Ferraro}, {Dalessandro},
  {Mucciarelli}, {Beccari}, {Miocchi}, {Bellazzini}, {Rich}, {Origlia},
  {Valenti}, {Rood}, \& {Ransom}}]{lanzoni10}
{Lanzoni}, B., {Ferraro}, F.~R., {Dalessandro}, E., {et~al.} 2010, \apj, 717,
  653

\bibitem[{{Lanzoni} {et~al.}(2013){Lanzoni}, {Mucciarelli}, {Origlia},
  {Bellazzini}, {Ferraro}, {Valenti}, {Miocchi}, {Dalessandro}, {Pallanca}, \&
  {Massari}}]{lanzoni13}
{Lanzoni}, B., {Mucciarelli}, A., {Origlia}, L., {et~al.} 2013, \apj, 769, 107

\bibitem[{{Libralato} {et~al.}(2022){Libralato}, {Bellini}, {Vesperini},
  {Piotto}, {Milone}, {van der Marel}, {Anderson}, {Aparicio}, {Barbuy},
  {Bedin}, {Borsato}, {Cassisi}, {Dalessandro}, {Ferraro}, {King}, {Lanzoni},
  {Nardiello}, {Ortolani}, {Sarajedini}, \& {Sohn}}]{libralato22}
{Libralato}, M., {Bellini}, A., {Vesperini}, E., {et~al.} 2022, \apj, 934, 150

\bibitem[{{Magg} {et~al.}(2022){Magg}, {Bergemann}, {Serenelli}, {Bautista},
  {Plez}, {Heiter}, {Gerber}, {Ludwig}, {Basu}, {Ferguson}, {Gallego},
  {Gamrath}, {Palmeri}, \& {Quinet}}]{magg_22}
{Magg}, E., {Bergemann}, M., {Serenelli}, A., {et~al.} 2022, \aap, 661, A140

\bibitem[{{Marigo} {et~al.}(2017){Marigo}, {Girardi}, {Bressan}, {Rosenfield},
  {Aringer}, {Chen}, {Dussin}, {Nanni}, {Pastorelli}, {Rodrigues}, {Trabucchi},
  {Bladh}, {Dalcanton}, {Groenewegen}, {Montalb{\'a}n}, \& {Wood}}]{marigo+17}
{Marigo}, P., {Girardi}, L., {Bressan}, A., {et~al.} 2017, \apj, 835, 77

\bibitem[{{Marino} {et~al.}(2009){Marino}, {Milone}, {Piotto}, {Villanova},
  {Bedin}, {Bellini}, \& {Renzini}}]{marino09}
{Marino}, A.~F., {Milone}, A.~P., {Piotto}, G., {et~al.} 2009, \aap, 505, 1099

\bibitem[{{Massari} {et~al.}(2023){Massari}, {Aguado-Agelet}, {Monelli},
  {Cassisi}, {Pancino}, {Saracino}, {Gallart}, {Ruiz-Lara},
  {Fern{\'a}ndez-Alvar}, {Surot}, {Stokholm}, {Salaris}, {Miglio}, \&
  {Ceccarelli}}]{massari+23}
{Massari}, D., {Aguado-Agelet}, F., {Monelli}, M., {et~al.} 2023, \aap, 680,
  A20

\bibitem[{{Massari} {et~al.}(2019){Massari}, {Koppelman}, \&
  {Helmi}}]{massari19}
{Massari}, D., {Koppelman}, H.~H., \& {Helmi}, A. 2019, \aap, 630, L4

\bibitem[{{Massari} {et~al.}(2012){Massari}, {Mucciarelli}, {Dalessandro},
  {Ferraro}, {Origlia}, {Lanzoni}, {Beccari}, {Rich}, {Valenti}, \&
  {Ransom}}]{massari12}
{Massari}, D., {Mucciarelli}, A., {Dalessandro}, E., {et~al.} 2012, \apjl, 755,
  L32

\bibitem[{{Massari} {et~al.}(2014){Massari}, {Mucciarelli}, {Ferraro},
  {Origlia}, {Rich}, {Lanzoni}, {Dalessandro}, {Valenti}, {Ibata}, {Lovisi},
  {Bellazzini}, \& {Reitzel}}]{Massari_14}
{Massari}, D., {Mucciarelli}, A., {Ferraro}, F.~R., {et~al.} 2014, \apj, 795,
  22

\bibitem[{{Mastrobuono-Battisti} {et~al.}(2019){Mastrobuono-Battisti},
  {Khoperskov}, {Di Matteo}, \& {Haywood}}]{mastrobuono19}
{Mastrobuono-Battisti}, A., {Khoperskov}, S., {Di Matteo}, P., \& {Haywood}, M.
  2019, \aap, 622, A86

\bibitem[{{Matteucci} \& {Brocato}(1990)}]{matteucci90}
{Matteucci}, F. \& {Brocato}, E. 1990, \apj, 365, 539

\bibitem[{{Matteucci} \& {Recchi}(2001)}]{matteucci_recchi2001}
{Matteucci}, F. \& {Recchi}, S. 2001, \apj, 558, 351

\bibitem[{{McKenzie} \& {Bekki}(2018)}]{mckenzie18}
{McKenzie}, M. \& {Bekki}, K. 2018, \mnras, 479, 3126

\bibitem[{{M{\'e}sz{\'a}ros} {et~al.}(2021){M{\'e}sz{\'a}ros}, {Masseron},
  {Fern{\'a}ndez-Trincado}, {Garc{\'\i}a-Hern{\'a}ndez}, {Szigeti}, {Cunha},
  {Shetrone}, {Smith}, {Beaton}, {Beers}, {Brownstein}, {Geisler}, {Hayes},
  {J{\"o}nsson}, {Lane}, {Majewski}, {Minniti}, {Munoz}, {Nitschelm},
  {Roman-Lopes}, \& {Zamora}}]{meszaros2021_omega}
{M{\'e}sz{\'a}ros}, S., {Masseron}, T., {Fern{\'a}ndez-Trincado}, J.~G.,
  {et~al.} 2021, \mnras, 505, 1645

\bibitem[{{Milone} {et~al.}(2017){Milone}, {Piotto}, {Renzini}, {Marino},
  {Bedin}, {Vesperini}, {D'Antona}, {Nardiello}, {Anderson}, {King}, {Yong},
  {Bellini}, {Aparicio}, {Barbuy}, {Brown}, {Cassisi}, {Ortolani}, {Salaris},
  {Sarajedini}, \& {van der Marel}}]{milone17}
{Milone}, A.~P., {Piotto}, G., {Renzini}, A., {et~al.} 2017, \mnras, 464, 3636

\bibitem[{{Minelli} {et~al.}(2021){Minelli}, {Mucciarelli}, {Massari},
  {Bellazzini}, {Romano}, \& {Ferraro}}]{minelli21}
{Minelli}, A., {Mucciarelli}, A., {Massari}, D., {et~al.} 2021, \apjl, 918, L32

\bibitem[{{Mucciarelli} {et~al.}(2012){Mucciarelli}, {Bellazzini}, {Ibata},
  {Merle}, {Chapman}, {Dalessandro}, \& {Sollima}}]{mucciarelli+12}
{Mucciarelli}, A., {Bellazzini}, M., {Ibata}, R., {et~al.} 2012, \mnras, 426,
  2889

\bibitem[{{Mucciarelli} {et~al.}(2015){Mucciarelli}, {Lapenna}, {Massari},
  {Pancino}, {Stetson}, {Ferraro}, {Lanzoni}, \& {Lardo}}]{mucciarelli15}
{Mucciarelli}, A., {Lapenna}, E., {Massari}, D., {et~al.} 2015, \apj, 809, 128

\bibitem[{{Mucciarelli} {et~al.}(2021){Mucciarelli}, {Massari}, {Minelli},
  {Romano}, {Bellazzini}, {Ferraro}, {Matteucci}, \& {Origlia}}]{mucciarelli21}
{Mucciarelli}, A., {Massari}, D., {Minelli}, A., {et~al.} 2021, Nature
  Astronomy, 5, 1247

\bibitem[{{Neumayer} {et~al.}(2020){Neumayer}, {Seth}, \&
  {B{\"o}ker}}]{neumayer2020}
{Neumayer}, N., {Seth}, A., \& {B{\"o}ker}, T. 2020, \aapr, 28, 4

\bibitem[{{Norris} {et~al.}(1996){Norris}, {Freeman}, \& {Mighell}}]{norris+96}
{Norris}, J.~E., {Freeman}, K.~C., \& {Mighell}, K.~J. 1996, \apj, 462, 241

\bibitem[{{Origlia} {et~al.}(2003){Origlia}, {Ferraro}, {Bellazzini}, \&
  {Pancino}}]{origlia+03}
{Origlia}, L., {Ferraro}, F.~R., {Bellazzini}, M., \& {Pancino}, E. 2003, \apj,
  591, 916

\bibitem[{{Origlia} {et~al.}(2025){Origlia}, {Ferraro}, {Fanelli},
  C.~{Lanzoni}, D., {Dalessandro}, \& {Pallanca}}]{origlia+25}
{Origlia}, L., {Ferraro}, F.~R., {Fanelli}, {et~al.} 2025, \aap

\bibitem[{{Origlia} {et~al.}(1997){Origlia}, {Ferraro}, {Fusi Pecci}, \&
  {Oliva}}]{origlia97}
{Origlia}, L., {Ferraro}, F.~R., {Fusi Pecci}, F., \& {Oliva}, E. 1997, \aap,
  321, 859

\bibitem[{{Origlia} {et~al.}(2013){Origlia}, {Massari}, {Rich}, {Mucciarelli},
  {Ferraro}, {Dalessandro}, \& {Lanzoni}}]{Origlia_13}
{Origlia}, L., {Massari}, D., {Rich}, R.~M., {et~al.} 2013, \apjl, 779, L5

\bibitem[{{Origlia} {et~al.}(2019){Origlia}, {Mucciarelli}, {Fiorentino},
  {Ferraro}, {Dalessandro}, {Lanzoni}, {Rich}, {Massari}, {Contreras Ramos}, \&
  {Matsunaga}}]{origlia_19}
{Origlia}, L., {Mucciarelli}, A., {Fiorentino}, G., {et~al.} 2019, \apj, 871,
  114

\bibitem[{{Origlia} \& {Rich}(2004)}]{origlia04}
{Origlia}, L. \& {Rich}, R.~M. 2004, \aj, 127, 3422

\bibitem[{{Origlia} {et~al.}(2002){Origlia}, {Rich}, \& {Castro}}]{origlia_02}
{Origlia}, L., {Rich}, R.~M., \& {Castro}, S. 2002, \aj, 123, 1559

\bibitem[{{Origlia} {et~al.}(2011){Origlia}, {Rich}, {Ferraro}, {Lanzoni},
  {Bellazzini}, {Dalessandro}, {Mucciarelli}, {Valenti}, \&
  {Beccari}}]{Origlia_11}
{Origlia}, L., {Rich}, R.~M., {Ferraro}, F.~R., {et~al.} 2011, \apjl, 726, L20

\bibitem[{{Origlia} {et~al.}(2005){Origlia}, {Valenti}, \& {Rich}}]{origlia05}
{Origlia}, L., {Valenti}, E., \& {Rich}, R.~M. 2005, \mnras, 356, 1276

\bibitem[{{Origlia} {et~al.}(2008){Origlia}, {Valenti}, \& {Rich}}]{origlia08}
{Origlia}, L., {Valenti}, E., \& {Rich}, R.~M. 2008, \mnras, 388, 1419

\bibitem[{{Palla}(2021)}]{palla2021}
{Palla}, M. 2021, \mnras, 503, 3216

\bibitem[{{Pallanca} {et~al.}(2021{\natexlab{a}}){Pallanca}, {Ferraro},
  {Lanzoni}, {Crociati}, {Saracino}, {Dalessandro}, {Origlia}, {Rich},
  {Valenti}, {Geisler}, {Mauro}, {Villanova}, {Moni Bidin}, \&
  {Beccari}}]{pallanca_21}
{Pallanca}, C., {Ferraro}, F.~R., {Lanzoni}, B., {et~al.} 2021{\natexlab{a}},
  \apj, 917, 92

\bibitem[{{Pallanca} {et~al.}(2021{\natexlab{b}}){Pallanca}, {Lanzoni},
  {Ferraro}, {Casagrande}, {Saracino}, {Purohith Bhaskar Bhat}, {Leanza},
  {Dalessandro}, \& {Vesperini}}]{pallanca21b}
{Pallanca}, C., {Lanzoni}, B., {Ferraro}, F.~R., {et~al.} 2021{\natexlab{b}},
  \apj, 913, 137

\bibitem[{{Pallanca} {et~al.}(2023){Pallanca}, {Leanza}, {Ferraro}, {Lanzoni},
  {Dalessandro}, {Cadelano}, {Vesperini}, {Origlia}, {Mucciarelli}, {Valenti},
  \& {Miola}}]{pallanca23}
{Pallanca}, C., {Leanza}, S., {Ferraro}, F.~R., {et~al.} 2023, \apj, 950, 138

\bibitem[{{Pancino} {et~al.}(2000){Pancino}, {Ferraro}, {Bellazzini}, {Piotto},
  \& {Zoccali}}]{pancino+00}
{Pancino}, E., {Ferraro}, F.~R., {Bellazzini}, M., {Piotto}, G., \& {Zoccali},
  M. 2000, \apjl, 534, L83

\bibitem[{{Pfeffer} {et~al.}(2021){Pfeffer}, {Lardo}, {Bastian}, {Saracino}, \&
  {Kamann}}]{pfeffer21}
{Pfeffer}, J., {Lardo}, C., {Bastian}, N., {Saracino}, S., \& {Kamann}, S.
  2021, \mnras, 500, 2514

\bibitem[{{Plez}(2012)}]{plez_12}
{Plez}, B. 2012, {Turbospectrum: Code for spectral synthesis}, Astrophysics
  Source Code Library, record ascl:1205.004

\bibitem[{{Romano} {et~al.}(2023){Romano}, {Ferraro}, {Origlia}, {Zwart},
  {Lanzoni}, {Crociati}, {Massari}, {Dalessandro}, {Mucciarelli}, {Rich},
  {Calura}, \& {Matteucci}}]{romano23}
{Romano}, D., {Ferraro}, F.~R., {Origlia}, L., {et~al.} 2023, \apj, 951, 85

\bibitem[{{Romano} {et~al.}(2010){Romano}, {Karakas}, {Tosi}, \&
  {Matteucci}}]{romano10}
{Romano}, D., {Karakas}, A.~I., {Tosi}, M., \& {Matteucci}, F. 2010, \aap, 522,
  A32

\bibitem[{{Romano} {et~al.}(2007){Romano}, {Matteucci}, {Tosi}, {Pancino},
  {Bellazzini}, {Ferraro}, {Limongi}, \& {Sollima}}]{romano+07}
{Romano}, D., {Matteucci}, F., {Tosi}, M., {et~al.} 2007, \mnras, 376, 405

\bibitem[{{Ryabchikova} \& {Pakhomov}(2015)}]{Ryabchikova_15}
{Ryabchikova}, T. \& {Pakhomov}, Y. 2015, Baltic Astronomy, 24, 453

\bibitem[{{Saracino} {et~al.}(2016){Saracino}, {Dalessandro}, {Ferraro},
  {Geisler}, {Mauro}, {Lanzoni}, {Origlia}, {Miocchi}, {Cohen}, {Villanova}, \&
  {Moni Bidin}}]{saracino16}
{Saracino}, S., {Dalessandro}, E., {Ferraro}, F.~R., {et~al.} 2016, \apj, 832,
  48

\bibitem[{{Saracino} {et~al.}(2019){Saracino}, {Dalessandro}, {Ferraro},
  {Lanzoni}, {Geisler}, {Cohen}, {Bellini}, {Vesperini}, {Salaris}, {Cassisi},
  {Pietrinferni}, {Origlia}, {Mauro}, {Villanova}, \& {Moni
  Bidin}}]{saracino19}
{Saracino}, S., {Dalessandro}, E., {Ferraro}, F.~R., {et~al.} 2019, \apj, 874,
  86

\bibitem[{{Saracino} {et~al.}(2015){Saracino}, {Dalessandro}, {Ferraro},
  {Lanzoni}, {Geisler}, {Mauro}, {Villanova}, {Moni Bidin}, {Miocchi}, \&
  {Massari}}]{saracino_15}
{Saracino}, S., {Dalessandro}, E., {Ferraro}, F.~R., {et~al.} 2015, \apj, 806,
  152

\bibitem[{{Tacchella} {et~al.}(2015){Tacchella}, {Lang}, {Carollo},
  {F{\"o}rster Schreiber}, {Renzini}, {Shapley}, {Wuyts}, {Cresci}, {Genzel},
  {Lilly}, {Mancini}, {Newman}, {Tacconi}, {Zamorani}, {Davies}, {Kurk}, \&
  {Pozzetti}}]{tacchella15}
{Tacchella}, S., {Lang}, P., {Carollo}, C.~M., {et~al.} 2015, \apj, 802, 101

\bibitem[{{Tolstoy} {et~al.}(2009){Tolstoy}, {Hill}, \& {Tosi}}]{tolstoy09}
{Tolstoy}, E., {Hill}, V., \& {Tosi}, M. 2009, \araa, 47, 371

\bibitem[{{Valenti} {et~al.}(2007){Valenti}, {Ferraro}, \&
  {Origlia}}]{valenti07}
{Valenti}, E., {Ferraro}, F.~R., \& {Origlia}, L. 2007, \aj, 133, 1287

\bibitem[{{Valenti} {et~al.}(2010){Valenti}, {Ferraro}, \&
  {Origlia}}]{valenti10}
{Valenti}, E., {Ferraro}, F.~R., \& {Origlia}, L. 2010, \mnras, 402, 1729

\bibitem[{{Valenti} {et~al.}(2005){Valenti}, {Origlia}, \&
  {Ferraro}}]{valenti05}
{Valenti}, E., {Origlia}, L., \& {Ferraro}, F.~R. 2005, \mnras, 361, 272

\bibitem[{{Weiland} {et~al.}(1994){Weiland}, {Arendt}, {Berriman}, {Dwek},
  {Freudenreich}, {Hauser}, {Kelsall}, {Lisse}, {Mitra}, {Moseley}, {Odegard},
  {Silverberg}, {Sodroski}, {Spiesman}, \& {Stemwedel}}]{weiland94}
{Weiland}, J.~L., {Arendt}, R.~G., {Berriman}, G.~B., {et~al.} 1994, \apjl,
  425, L81

\bibitem[{{White} \& {Rees}(1978)}]{white_rees1978}
{White}, S.~D.~M. \& {Rees}, M.~J. 1978, \mnras, 183, 341

\end{thebibliography}
%\begin{thebibliography}{}
%\bibitem[Valenti et al.(2010)]{valenti_10} Valenti, E., Ferraro, F.~R., \& Origlia, L.\ 2010, \mnras, 402, 1729. doi:10.1111/j.1365-2966.2009.15991.x
%\bibitem[Origlia et al.(2003)]{origlia_03} Origlia, L., Ferraro, F.~R., Bellazzini, M., et al.\ 2003, \apj, 591, 916. doi:10.1086/375363
%\end{thebibliography}

%panel con 
\end{document}